\documentclass[sigconf]{acmart}

\AtBeginDocument{%
	\providecommand\BibTeX{{%
			\normalfont B\kern-0.5em{\scshape i\kern-0.25em b}\kern-0.8em\TeX}}}
		
\setcopyright{acmcopyright}
\copyrightyear{2023}
\acmYear{2023}
\acmDOI{XXXXXXX.XXXXXXX}
		
\acmConference[ASIA CCS'24]{Proceedings of the 2024 ACM ASIA Conference on Computer and Communications Security}{July 1--5, 2024}{Singapore.}
\acmISBN{978-1-4503-XXXX-X/18/06}
		
\usepackage{amsfonts}
\usepackage{algorithm}
\usepackage[noend]{algorithmic}
\usepackage[utf8]{inputenc}
\usepackage{multirow}
\usepackage{textcomp}
\usepackage{xcolor,colortbl,siunitx}
\def\BibTeX{{\rm B\kern-.05em{\sc i\kern-.025em b}\kern-.08em
    T\kern-.1667em\lower.7ex\hbox{E}\kern-.125emX}}
\usepackage{url}
\usepackage{soul}

\usepackage{tikz}

\newcommand{\refined}{{}^\star}

\newcommand{\clang}{\textsc{clang}}
\newcommand{\lfence}{\textsc{lfence}}
\newcommand{\csdb}{\textsc{csdb}}
\newcommand{\csel}{\textsc{csel}}

\newcommand{\dmb}{\textsc{dmb}}
\newcommand{\dsb}{\textsc{dsb}}
\newcommand{\isb}{\textsc{isb}}

\newcommand{\slh}{\textsc{slh}}
\newcommand{\angr}{\textit{angr}}
\newcommand{\scamv}{Scam-V}
\newcommand{\mstates}{S}
\newcommand{\mstate}{s}

\newcommand{\obsmodel}[1]{\mathit{M}_{#1}}
\newcommand{\Obs}{\mathit{O}}
\newcommand{\obs}[1]{\mathit{o}_{#1}}

\newcommand{\bir}{BIR}
\newcommand{\vex}{VEX}
\newcommand{\regname}[1]{\texttt{\textbf{#1}}}
\newcommand{\projname}{\pi}
\newcommand{\projection}[2]{\projname(#1) = #2}
\newcommand{\optzero}{\textbf{-O0}}
\newcommand{\opttwo}{\textbf{-O2}}

\newcommand\algorithmicprocedure{\textbf{procedure}}
\newcommand{\algorithmicendprocedure}{\algorithmicend\ \algorithmicprocedure}
\makeatletter
\newcommand\PROCEDURE[3][default]{%
  \ALC@it
  \algorithmicprocedure\ \textsc{#2}(#3)%
  \ALC@com{#1}%
  \begin{ALC@prc}%
}
\newcommand\ENDPROCEDURE{%
  \end{ALC@prc}%
  \ifthenelse{\boolean{ALC@noend}}{}{%
    \ALC@it\algorithmicendprocedure
  }%
}
\newenvironment{ALC@prc}{\begin{ALC@g}}{\end{ALC@g}}
\makeatother

\begin{document}

\title[Beyond Over-Protection]{Beyond Over-Protection: A Targeted Approach to Spectre Mitigation and Performance Optimization}

\author{Tiziano Marinaro}
\affiliation{%
  \institution{CISPA Helmholtz Center for}
  \streetaddress{}
  \city{}
  \country{}}
\affiliation{%
  \mbox{\institution{Information Security, Saarland University}}
  \streetaddress{}
  \city{}
  \country{}}
\email{tiziano.marinaro@cispa.de}

\author{Pablo Buiras}
\affiliation{%
  \institution{KTH Royal Institute of Technology}
  \streetaddress{}
  \city{}
  \country{}}
\email{buiras@kth.se}

\author{Andreas Lindner}
\affiliation{%
  \institution{KTH Royal Institute of Technology}
  \streetaddress{}
  \city{}
  \country{}}
\email{andili@kth.se}

\author{Roberto Guanciale}
\affiliation{%
  \institution{KTH Royal Institute of Technology}
  \streetaddress{}
  \city{}
  \country{}}
\email{robertog@kth.se}

\author{Hamed Nemati}
\affiliation{%
  \institution{KTH Royal Institute of Technology}
  \streetaddress{}
  \city{}
  \country{}}
\email{hnnemati@kth.se}

\begin{abstract}
Since the advent of Spectre attacks, researchers and practitioners have developed a range of hardware and software measures to counter transient execution attacks. A prime example of such mitigation is \emph{speculative load hardening} in LLVM, which protects against leaks by tracking the speculation state and masking values during misspeculation. LLVM relies on static analysis to harden programs using \slh{} that often results in over-protection, which incurs performance overhead.
We extended an existing side-channel model validation framework, \scamv{}, to check the vulnerability of programs to Spectre-PHT attacks and optimize the protection of programs using the \slh{} approach. We illustrate the efficacy of \scamv{} by first demonstrating that it can automatically identify Spectre vulnerabilities in real programs, e.g., fragments of crypto-libraries. We then develop an optimization mechanism that validates the necessity of \slh{} hardening w.r.t. the target platform. Our experiments showed that hardening introduced by LLVM in most cases could be significantly improved when the underlying microarchitecture properties are considered.
\end{abstract}

\keywords{hardware security, side-channel attacks, countermeasures, Spectre}

\begin{CCSXML}
<ccs2012>
   <concept>
       <concept_id>10002978.10003001.10010777.10011702</concept_id>
       <concept_desc>Security and privacy~Side-channel analysis and countermeasures</concept_desc>
       <concept_significance>300</concept_significance>
       </concept>
 </ccs2012>
\end{CCSXML}

\ccsdesc[300]{Security and privacy~Side-channel analysis and countermeasures}

\maketitle

\section{Introduction}

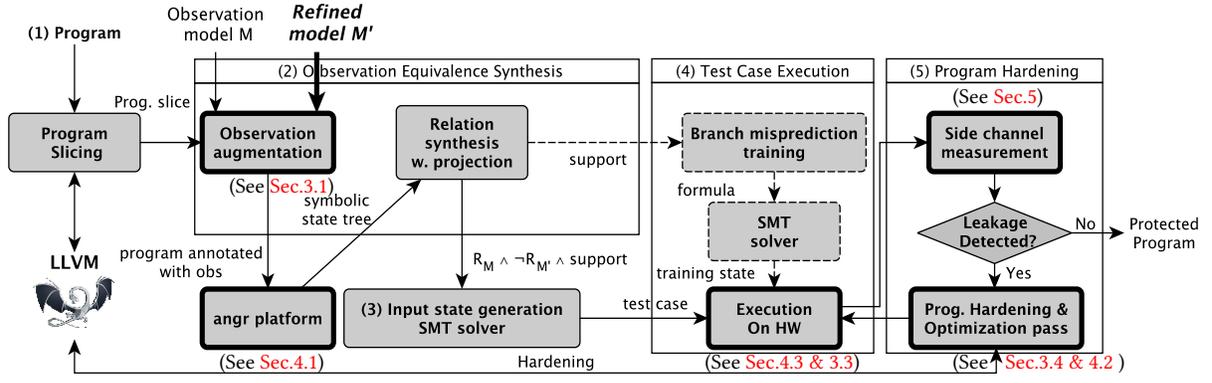
\begin{figure*}[t]
\tikzset{every picture/.style={line width=0.7pt}}
\begin{tikzpicture}[x=0.75pt,y=0.75pt,yscale=-1,xscale=1]

\draw (350,117) node  {\includegraphics[width=450pt,height=140pt]{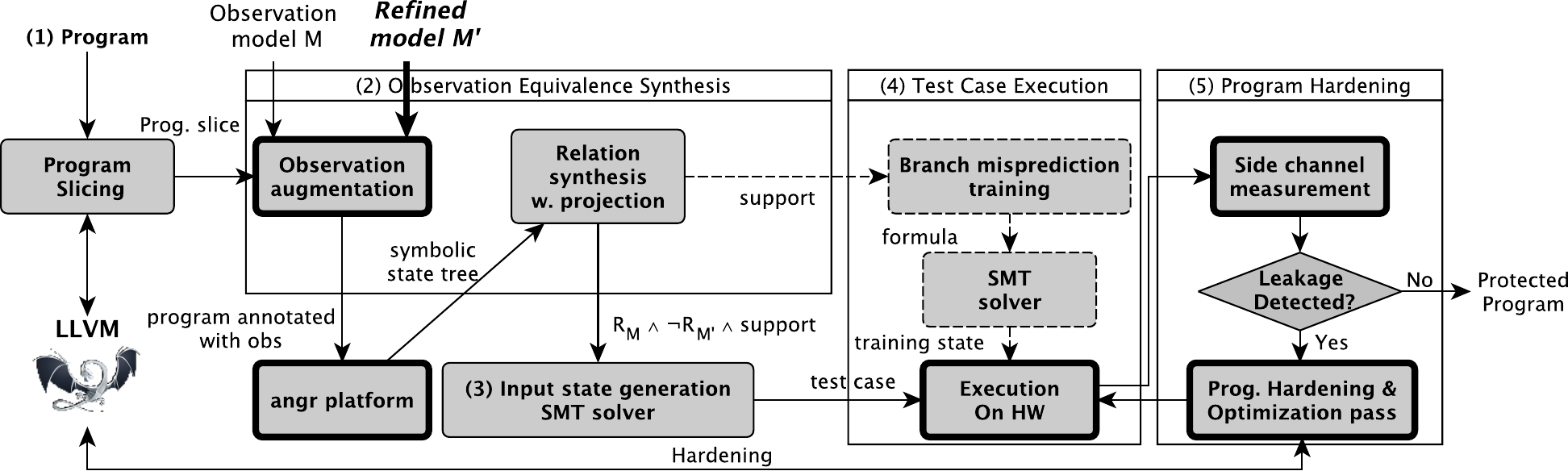}};
% Text Node
\draw (155,199) node [anchor=north west][inner sep=0.75pt]  [font=\small] [align=left] {{\small  (See \textcolor{red}{Sec.\ref{subsec:angrintegration}})}};
% Text Node
\draw (160,110) node [anchor=north west][inner sep=0.75pt]  [font=\small] [align=left] {{\small  (See \textcolor{red}{Sec.\ref{subsec:obsref}})}};
% Text Node
\draw (400,199) node [anchor=north west][inner sep=0.75pt]  [font=\small] [align=left] {{\small  (See \textcolor{red}{Sec.\ref{para:experiment-setup} \& \ref{sec:micro-state-config}})}};
% Text Node
\draw (525,199) node [anchor=north west][inner sep=0.75pt]  [font=\small] [align=left] {{\small  (See\;\; \textcolor{red}{Sec.\ref{subsec:fenceplacement} \&~\ref{subsec:fenceplacementimpl}} })};
% Text Node
\draw (525,65) node [anchor=north west][inner sep=0.75pt]  [font=\small] [align=left] {{\small  (See \textcolor{red}{Sec.\ref{sec:evaluation}})}};

\end{tikzpicture}
  \caption{\scamv{'s} workflow.  Modules in bold boxes are those that have been added or modified in this paper.}
  \label{fig:scamv_arch}
  \Description{Workflow of Scam-V pipeline. Step 1: program slicing, step 2: observation augmentation, step 3: symbolic execution in angr, step 4: relation synthesis with projection, step 5: input state generation by SMT solver, step 6: execution on hardware, step 7: side channel measurement, step 8: check the leakage, step 9: program hardening, step 10: hardening optimization}
\end{figure*}

The past decade has witnessed a surge in the number of microarchitectural attacks that exploit hardware side channels to exfiltrate secret information.
The prime example of such attacks is the Spectre attack family~\cite{spectre} which leverages transient (speculative) execution to leak data through caches.
To counter transient execution vulnerabilities, researchers and practitioners have developed several analysis techniques as well as hardware and software measures~\cite{sok:hwmitigations,cauligi:sok-spectre-sw-defense}.
Examples include \emph{fence instructions}, like x86 \lfence{}, which are \emph{serializing instructions} to mitigate the side effects of speculative execution. Another example of software-based mitigation is \emph{Speculative Load Hardening} (\slh{})~\cite{Carruth2020}, which protects against transient execution leakages by tracking the speculation state and masking values during misspeculation. 

Unfortunately, to date, almost all existing mitigations against transient execution leakages are either incomplete and miss attacks or overly conservative and slow~\cite{cauligi:sok-spectre-sw-defense}.  For example,
the only mitigation that guarantees complete protection against the Spectre-V1 (-PHT) attack on commodity processors are memory fences, such as  \lfence{}, \csdb{} (used in the implementation of the \slh{} pass for AArch64), and \dsb{}+\isb{}. Nevertheless, using too many fences (over-fencing) hinders performance (e.g., inserting a fence at every load or control flow point can incur around 440\% overhead~\cite{Oleksenko2018}), while using too few fences (under-fencing) may allow unexpected leakage to occur.
There are also tools such as LLVM~\cite{LattnerAdve:llvm} that harden programs, e.g., using \slh{}, against transient leakages. However, almost all existing software-based solutions rely on \textbf{static} analysis and follow a conservative approach to harden programs and thus suffer from over-protection. Moreover, different processors implementing the same architecture can have substantially different speculative leakage. This suggests we may need to rethink the way we analyze and protect programs against transient attacks.

\textbf{Our Approach.}\ In this paper, we present a systematic approach to identifying conditions (code alignment and compiler optimization level) under which programs become vulnerable to Spectre-PHT attacks and finding an optimal hardening scheme to protect programs against potential leakages. Our approach relies on relational testing w.r.t the underlying processor's real implementation to determine the necessity of the hardening applied by state-of-the-art approaches like the LLVM compiler infrastructure.

We build on an existing platform, \scamv{}~\cite{scamv,scamv:buiras2021}, which automates the validation of abstract side-channel models via relational testing. Fig.~\ref{fig:scamv_arch} shows the \scamv{'s} workflow. 
The choice of \scamv{} is supported by the fact that compared to existing tools like Revizor~\cite{revizor}, \scamv{} requires fewer test cases to disclose leakages due to its testing approach and internal optimizations (see~\ref{sec:obsref}).
\scamv{}, first, generates test cases consisting of a program and two input states that are in an equivalence relation that is automatically synthesized w.r.t model under test. Then, \scamv{} executes the test cases on real hardware and measures the side channel to find cases that invalidate the input model.

The current implementation of \scamv{} is insufficient for our purposes, and several changes were required to enable \scamv{} to test the information-flow security of real-world programs, e.g., fragments of OpenSSL (Sec.~\ref{sec:evaluation}). Sections~\ref{sec:method} and~\ref{sec:implementation} describe our methodology and changes to the \scamv{'s} pipeline. Particularly, instead of testing random binaries, we feed \scamv{} with slices taken from the binary of programs under test. Moreover, \scamv{} uses symbolic execution to synthesize the equivalence relation for the input states. Nevertheless, the existing simple symbolic execution engine in \scamv{} could not handle  slices taken from real programs. Therefore, we further connect \scamv{} to the \emph{angr} framework~\cite{angr}; Sec~\ref{sec:implementation} elaborates on the details of this integration. Also, we have generalized the \emph{refinement} technique that \scamv{} uses to generate inputs that are more likely to trigger leakages~\cite{scamv:buiras2021}. Finally, we have connected \scamv{} to LLVM and developed a compiler pass to optimally protect programs against leakages.

\textbf{Results.}\ Our experiments highlight several interesting results, showing that (i) unexpected microarchitectural details like the \textbf{alignment}\footnote{Alignment refers to the process of arranging the binary instructions in memory.} of the code executing on the processor can affect the leakage behavior of programs; (ii) most hardening applied by the current approaches are not required in real executions; and (iii) there are transient leakages like the one presented in~\cite{scamv:buiras2021} that are not mitigated by the implementation of existing mitigations such as LLVM AArch64 \slh{}\footnote{This case has been theoretically proved~\cite{patrignani2021exorcising} before.}.

LLVM \textbf{conservatively} hardens programs to stop transient execution leakages. For example, in LLVM AArch64 \slh{} pass, almost every load instruction is hardened. However, our experiments show that hardening programs to stop potential leakages strictly depends on the hardware platform that programs execute on. 
That is, a program that leaks on a specific architecture or processor does not necessarily show the same behavior when it executes on a different architecture or core. This behavior is witnessed by a few cases from the Kocher Spectre benchmarks, e.g., Case \#10~\cite{kocher:sepcterv1-benchmarks}. While all existing analysis tools and techniques, e.g.,~\cite{guarnieri:contracts,ctfoundations:pldi:20,lcms:mosier:2022,2021:huntingthehaunter,2021:oo7}, classify this case as a leaky program and conservatively protect it using fences, we verified that this test case does not induce any leakages on ARM processors we have tested (Cortex-A53 and -A72), and therefore, there is no need for any protection on these cores.
Based on this insight, we have built a toolchain to examine the vulnerability of programs w.r.t the target hardware platform, and optimally harden them to stop the found leakages.

\section{Background}

\subsection{Side channels}
\label{sec:background-spectre}
Side channels are unintended information flow paths that are potentially exploitable by a malicious process to exfiltrate secrets from the memory of other processes.
A prime example of side-channel vulnerabilities is \textit{Spectre attacks}. These attacks are characterized by a \textit{speculation primitive}~\cite{microsoft:taxonomy}, which enables an instruction that accesses a secret to supply it to a transient \textit{transmitter}~\cite{jiyong:stt} that leaks it.
Speculation primitive is some hardware mechanism that initiates speculative execution.  Examples include control-flow instructions (e.g., conditional or indirect branches~\cite{spectre}, return instructions~\cite{spectrersb}), which predict the intended value of the program counter and loads which predict the effective addresses of prior
unresolved stores~\cite{spectrev4}.

Spectre attacks usually leak data via data caches, which can be measured by an attacker using techniques such \emph{Flush+Reload}~\cite{flushreload}.
In Flush+Reload, the attacker flushes cache entries, allowing the victim to (normally or speculatively) execute a secret-dependent load, after which the attacker measures the reload time. This reloads time enables the attacker to determine if the victim has accessed any cache entries during its execution and subsequently enables the attacker to extract the secret.

\begin{figure*}[t]
\centering
\includegraphics[width=0.9\linewidth]{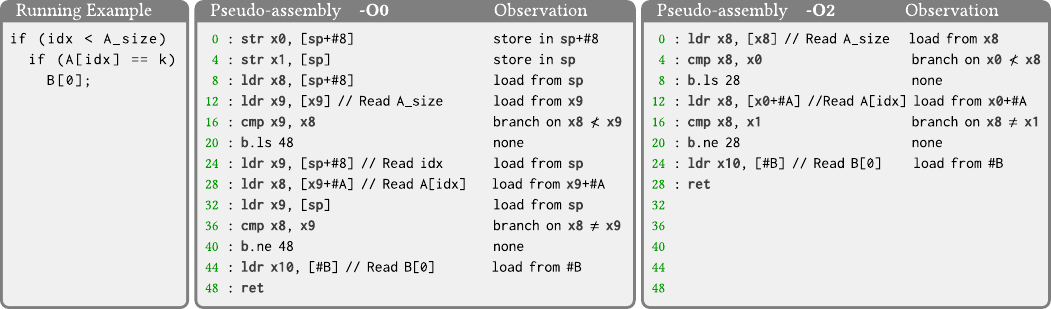}
\caption{The running example. Benchmark Case \#10 from~\cite{kocher:sepcterv1-benchmarks}.}
\label{fig:running_example}
\Description{The running example in C code, in pseudo-assembly with no optimization and in pseudo-assembly with optimization level O2.}
\end{figure*}

\textbf{Running example.} We introduce the example of Fig.~\ref{fig:running_example} to show the main concepts of our approach. The example is a simplified version of Case \#10 from the benchmarks used in~\cite{kocher:sepcterv1-benchmarks,guarnieri:spectector,ctfoundations:pldi:20}. The program consists of two nested conditional statements, where the first one checks the bound of an accessed index within the public array $A$. The load of the first element from the array $B$ is then decided by the inner \texttt{if} statement by checking whether or not the accessed value in $A$ based on the provided index is equal to $k$.

When the processor encounters the first \texttt{if} statement, based on its internal state and the pipeline load, it may decide to speculatively execute the read from the array $A$ even when $idx$ is greater or equal to $A\_size$. This potentially enables an attacker who controls the value of $idx$ and $k$ to get access to sensitive information inferred from the comparison in the inner condition. Specifically, the attacker can infer the value of data loaded in speculation by executing $A[idx]$ when the $B[0]$ shows up in the cache, indicating the condition in the inner \texttt{if} statement holds. 

When analyzing the security of a program, it is essential to consider the security level of variables to get a precise understanding of the program leakage potential and decide the specific mitigation which must be applied. For instance,  depending on whether $idx$ is deemed private or public, our running example leaks differently. While in the latter case, the leakage happens when the execution reaches $B[0]$, i.e., the inner conditional expression holds, the former case shows different leakage at the point $A[idx]$ is executed.

\subsection{Mitigations of transient execution attacks}
Spectre attacks can be mitigated through a combination of hardware and software measures. Using \emph{fence} instructions or \emph{speculative load hardening} are examples of such measures.

\paragraph{Fence instructions.} 
To mitigate Spectre-PHT, ARM architecture includes different fences like \dmb{} (data memory barrier), \dsb{} (data synchronization barrier), and \isb{} (instruction synchronization barrier) instructions.
Another AArch64 barrier, which is used by LLVM's \slh{} pass, is \csdb{} or \emph{Consumption of Speculative Data Barrier}. \csdb{} restricts speculative execution and data value prediction. Once  \csdb{} is executed, no instruction other than branch instructions in program order can be speculatively executed using the results of data value predictions or \textsc{PSTATE} predictions of any instructions appearing before \csdb{} that have not been resolved. This allows for control flow speculation before and after \csdb{} and permits speculative execution of conditional data processing instructions after \csdb{}, as long as they don't use the results of predictions made before \csdb{}.

\paragraph{Speculative load hardening} An alternative approach is \emph{speculative load hardening} (\slh{})~\cite{Carruth2020}, which masks addresses or values of loads inside in a conditional branch with the branch predicate. The idea of \slh{} is to maintain a predicate indicating if the execution is currently in a mispredicted branch or not. This predicate is then used to “poison” either the outputs (i.e., values) or inputs (i.e., addresses) of load instructions. 
Since \slh{} limits the amount of speculative execution that can be performed, it also impacts the programs' performance. However, compared to other mitigation, \slh{} is more efficient, with a speed improvement of approximately 1.77 times. The overhead of load hardening is expected to range from 10\% to 50\%, with most large applications experiencing an overhead of $\sim$30\%~\cite{slh}.
\paragraph{LLVM AArch64 \slh}
The LLVM \slh{} pass for AArch64 uses taint tracking (blue lines) to find program points that must be protected.

\begin{figure}[h]
\centering
\includegraphics[width=0.9\linewidth]{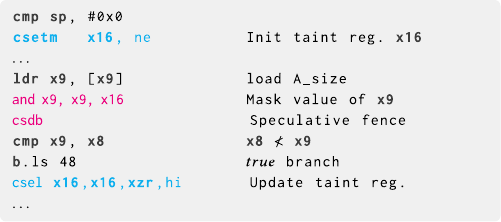}
\Description{Example of Speculative load hardening performed by the LLVM Pass on the pseudo-assembly of the running example with no optimization.}
\end{figure}

The snippet shows the hardened version of our example using the LLVM \slh{} pass. AArch64 \slh{} reserves the register \regname{x16} as the taint register, which contains all-ones when no misspeculation happens, and all-zeros when misspeculation is detected. Mask operations (red lines) are inserted using the \textsc{and} instruction, and misspeculation is tracked using a \emph{data-flow conditional select instruction} (\csel{}) based on the evaluation of CPU flags, which can be limited with \csdb{} to avoid its speculation.
When a conditional branch direction gets misspeculated, the semantics of the inserted \csel{} instruction is such that the taint register will contain all zero bits.

\paragraph{AArch64 \slh{} limitation}
\label{background:aarch64-slh-limitation}Current implementation of the AArch64 \slh{} can only protect against data leak through memory, and it does not prevent leakages through registers. An example of the latter case is when $idx$ in our running example is labelled as private and the memory load at line 12 in the optimized (\opttwo{}) assembly code of Fig~\ref{fig:running_example} leaks the value of $idx$.

Besides, \slh{} conservatively hardens programs, and some of the introduced hardenings can safely be removed without affecting the security of the hardened code. For example, in case $idx$ is a public value, the hardening at line 24 in the unoptimized (\optzero{}) assembly code of Fig~\ref{fig:running_example} is unnecessary and can be safely removed. Optimizing the \slh{} hardening can, in some cases, be done using static analysis, e.g., it can detect that the hardening at 24 is not necessary when $idx$ is public. Nevertheless, static analysis can not always be helpful and full optimization requires taking into account the properties of the underlying microarchitecture and must be guided by analyzing the execution traces of programs on real hardware.

\subsection{Side channel analysis --- \scamv{}}
In the following, we use $\mstate \in \mstates$ to range over ISA states and we say that two states $\mstate_1$ and $\mstate_2$ are \emph{low-equivalent} (i.e., $\mstate_1 =_{L}\mstate_2$) iff
they agree on every public register and memory location.
We also say that two states $\mstate_1$ and $\mstate_2$ are \emph{indistinguishable} iff an attacker cannot distinguish executions on real hardware that start from the same microarchitectural state and any states corresponding to $\mstate_1$ and $\mstate_2$.
Intuitively, a system is free of side channels if low-equivalent states are also indistinguishable. That is,
the attacker cannot learn anything from the channel that is not already public.

Since verification requires models, verifying the absence of leakage due to side channels requires a model capturing the channels. However, due to the complexity of modern processors, it is infeasible to explicitly model all the complex and intertwined microarchitectural features like cache hierarchies, cache replacement policies, as well as memory and buses. Abstract \textit{observational models}, hereafter denoted by $\obsmodel{i}$, tackle this problem by overapproximating attacker capabilities. To this end, the abstract model of the system is extended with a set of possible observations $\obs{} \in \Obs{}$, e.g., cache tag or set index,  and a transition relation $\rightarrow \subseteq \mstates \times \Obs \times \mstates$. For each such model, the observations represent the part of the processor state that may affect the channel at each transition.

For instance, in order to overapproximate the information leakage that may occur in Fig.~\ref{fig:running_example} due to the presence of caches, the processor model may be extended with the observations that are shown in the right column: e.g., the execution of the first instruction of the non-optimised binary may leak the value of stack pointer plus 8. Intuitively, this model, say $\obsmodel{}$, captures that the program execution time depends on the addresses of loads and instructions that are executed based on conditional branches.
We say that two states $\mstate,\mstate' \in \mstates$
are \emph{observationally equivalent} (i.e., $\mstate \sim_M
\mstate'$), iff for every possible trace $\mstate \xrightarrow{\obs{1}} \mstate_1 \dots
\xrightarrow{\obs{n}} \mstate_n$ of $M$ there
is a trace $\mstate' \xrightarrow{\obs{1}'} \mstate'_1 \dots \xrightarrow{\obs{n}'} \mstate'_{n}$ such that  $[\obs{1}, \dots, \obs{n}] = [\obs{1}', \dots, \obs{n}']$.

Clearly, we would like models to be  \emph{sound}, which means that observationally equivalent states should lead to executions that
cannot be distinguished by an attacker on real hardware.

\begin{definition}[Soundness]\label{def:obsmodelsoundness}
  An observational model $M$ is \emph{sound} if  $\mstate \sim_M
\mstate'$ entails indistinguishability of $\mstate$ and $\mstate'$.
\end{definition}
Sound observational models can be regarded as reliable foundations for side-channel analysis, since they allow to demonstrate the absence of side channels by statically proving that low-equivalent states are observationally equivalent.
The problem with speculative execution is that observational models that do not take into account transient observations
are unsound, and there is no general model of transient observations that covers all possible microarchitectures.

\scamv{}~\cite{scamv-tool,scamv,scamv:buiras2021} combines techniques from program verification and fuzzing to perform \emph{relational testing} and examine the soundness of observation models. At a high level (see Fig.~\ref{fig:scamv_arch}), (1) it generates well-formed binary programs and (2) synthesizes a relation that, for a generated program, identifies states that are observationally equivalent according to the model under the validation. Then, (3) an instance of this relation in terms of two input states is generated. Finally, (4) \scamv{} runs the generated binary with different input pairs on real hardware and compares the measurements on the side channel of the real processor. Since the generated inputs satisfy the synthesized relation, the \emph{soundness} of the model would imply that the side-channel data on hardware cannot be distinguished either. Thus, a test case where we can distinguish the two runs on the hardware amounts to a counterexample (a potential side channel) that invalidates the observational model.

\newcommand{\symsymb}[1]{\alpha_{#1}}
\scamv{} uses the symbolic execution of programs that are annotated with observations to synthesize the equivalence relation. Symbolic execution is a technique to explore all program execution paths by using symbolic values instead of concrete ones for the inputs.
Starting from an initial symbolic state, the execution explores all possible paths and collects the execution effects in a final symbolic state $\sigma \in \Sigma$ for each path.
Each symbolic state $\sigma$ consists of a map from variables to symbolic expressions (i.e., where symbols represent initial state variables) and a path condition $p_{\sigma}$ (i.e., a symbolic expression identifying the condition that leads to the execution of that path). \scamv{} also maintains a list of symbolic expressions $l_{\sigma}$, which collects the
effects of observable statements that have been encountered. E.g., starting from an initial state which maps each register $x_{i}$ to a symbol $\symsymb{i}$ and the memory to $\symsymb{M}$, an empty list of observations, and a $true$ path condition, the symbolic execution of
the \opttwo{} program of Fig.~\ref{fig:running_example} produces three final states:
\begin{itemize}
  \item one for $idx \geq A\_size$, where path condition $\symsymb{0} \geq \symsymb{M}[\symsymb{8}]$, state mapping $x_{8} \mapsto \symsymb{M}[\symsymb{8}]$, and observation list $[\symsymb{8};false]$,
  \item one for $idx < A\_size$ and $A[idx] \neq k$,
        where path condition $\symsymb{0} < \symsymb{M}[\symsymb{8}] \land  \symsymb{M}[\symsymb{0} + \#A] \neq \symsymb{1}$, state mapping $x_{8} \mapsto \symsymb{M}[\symsymb{0}]$, and observation list $[\symsymb{8}; true; \symsymb{0}+\#A; false$],
 \item one for $idx < A\_size$ and $A[idx] = k$,
        where path condition $\symsymb{0} < \symsymb{M}[\symsymb{8}] \land  \symsymb{M}[\symsymb{0} + \#A] \neq \symsymb{1}$, state mapping $x_{8} \mapsto \symsymb{M}[\symsymb{0}+\#A], x_{10} \mapsto \symsymb{M}[\#B]$, and observation list $[\symsymb{8}; true; \symsymb{0}+\#A; true; \#B]$.
\end{itemize}

\scamv{} uses self composition~\cite{selfcomposition} to compute the observational equivalence relation by imposing equivalence of the symbolic observation list of the final states of the symbolic execution. For the example above, the relation would be as follows:
\[
  \begin{array}{ll}
    &
  \symsymb{8} = \symsymb{8}' \land (\symsymb{0} \geq \symsymb{M}[\symsymb{8}] \Leftrightarrow \symsymb{0}' \geq \symsymb{M}'[\symsymb{8}']) \land \\
  (\symsymb{0} < \symsymb{M}[\symsymb{8}] \Rightarrow  & \symsymb{0} = \symsymb{0}' \land \\
&
  \symsymb{M}[\symsymb{0} + \#A] \neq \symsymb{1}
  \Leftrightarrow
  \symsymb{M}'[\symsymb{0}' + \#A] \neq \symsymb{1}')
  \end{array}
\]

\subsubsection{Observation Models Refinement}\label{sec:obsref}
Using relational testing similar to the \scamv{'s} approach to validate observational models can lead to state space explosion. This is because an unguided search may explore states that are either too similar to each other, thus unlikely to invalidate the given model or fail to trigger attacker-visible microarchitectural behavior. \emph{Observation refinement}~\cite{scamv:buiras2021} tackles this problem by adding more fine-grained observations of the system state to capture behaviors we need to exclude. 

For a given program, the observation model $\obsmodel{}$ partitions the input states into observation equivalence classes. Relevant pairs in such equivalence classes must be tested to validate the soundness of $\obsmodel{}$. To make validation more efficient, the observation refinement suggests further repartitioning the induced equivalence classes using a complementary model $\obsmodel{}'$ that captures the observations that might arise from the side channel under scrutiny.

For instance, in ARM Cortex-M0, a 32-bit multiplication instruction can take between 1 and 32 clock cycles to complete, depending on the operands being multiplied~\cite{cortexm0}. 
The number of clock cycles required to execute a multiplication operation can potentially leak information about the operands being multiplied, revealing sensitive information about the cryptographic algorithm being executed.
Arithmetic operations like multiplication normally are not assigned an observation in standard observational models. However, to check for the existence of the side channel above,  one can define a refined observation model which reveals the most significant bits of the multiplication operator:
\begin{figure}[h]
\centering
\includegraphics[width=0.95\linewidth]{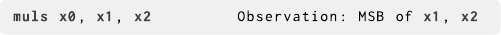}
\vspace{-0.5em}
\Description{Example of a refined observation of a multiplication instruction that defines as public the most significant bits of its operators contained in registers x1 and x2.}
\end{figure}

Having defined a refined observation model, test cases (pairs of states $\mstate$ and $\mstate'$) are chosen s.t. they are observationally equivalent w.r.t $\obsmodel{}$ yet are distinguishable w.r.t $\obsmodel{}'$, i.e., $\mstate{} \sim_{\obsmodel{} \wedge \neg \obsmodel{}'} \mstate{}'$.
Observation refinement is especially important to facilitate finding transient execution leakages by steering the test case generation toward generating input states that ensure distinguishable cache updates due to misspeculated memory load instructions.

%%% Local Variables:
%%% mode: latex
%%% TeX-master: "../main"
%%% End:

\section{Methodology}
\label{sec:method}

To efficiently protect against microarchitectural leakages, it's essential to design a mitigation strategy to ensure that the resulting system meets both the necessary security and performance requirements. In doing that, an obstacle is the lack of documentation or model regarding the speculative leakage of different microarchitectures.
We address this problem by using \emph{relational testing} to identify when a specific program is affected by transient leakages on a specific microarchitecture. This allows us to remove unnecessary mitigation and protect only the parts of the program that are indeed vulnerable on a given processor. The general strategy consists of three tasks: (a) use \scamv{} to generate pairs of input states that should trigger potential information leakages in case of speculative behavior; (b) test if the leakage indeed occurs; (c) protect the vulnerable program fragments and remove unnecessary hardening.

\begin{figure}[]
\centering
\includegraphics[width=0.9\linewidth]{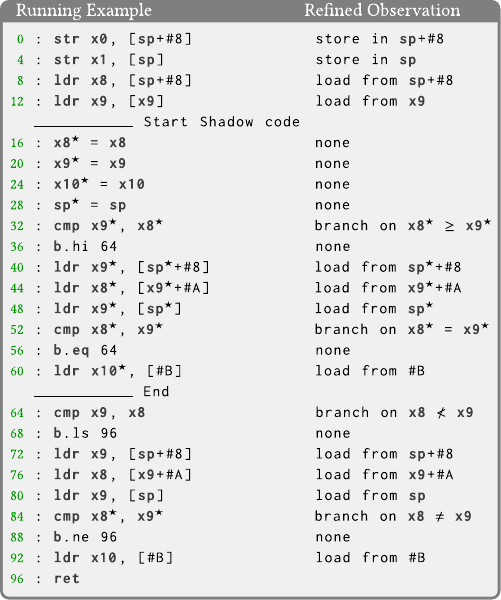}
\caption{The running example instrumented with shadow code and refined observations.}
\label{fig:running:eg:shadow:code}
\Description{The running example in pseudo-assembly with no optimization instrumented with shadow code and refined observations.}
\end{figure}

\subsection{Observation refinement for speculation}
\label{subsec:obsref}
In order to drive the generation of test cases, we leverage \emph{observation refinement} (see Sec.~\ref{sec:obsref}). However, the previous implementation of the refinement technique in \scamv{} could only handle a single conditional branch and was insufficient for our tests. Thus, we generalized the observation refinement in \scamv{} by introducing a new program transformation that simulates the speculative execution of program instructions.
This refinement enables us to observe memory operations that happen in speculation.

Our experiments focus on branch prediction, but the technique is general enough to cover other types of speculation.
We achieve refinement by choosing a set of branches that can be misspeculated and composing them with the program. This is in line with the Spectector approach, which uses the always misprediction policy~\cite{guarnieri:spectector}.

Our program transformation inlines a shadow copy of the program fragment starting from the execution point before the first misspeculation.
We statically fix the size of these fragments to cover the largest expected speculative window of the processor.
This shadow code (marked with $\refined$ in Fig.~\ref{fig:running:eg:shadow:code}) is a copy of the original program fragment, where all selected misspeculating branches have negated conditions---to simulate what can happen in misspeculation. The shadow code also operates over a shadow machine state in order to not affect the non-speculative behavior.
All memory operations executed in the shadow code raise a refined observation, indicating that they are possible causes of leakage.

Fig.~\ref{fig:running:eg:shadow:code} shows how this transformation works for our running example.
In this example, we inline a shadow copy (code snippet between \texttt{Start} and \texttt{End}) of the program fragment starting from the `\texttt{if}' statement with the negated condition. During the execution, we save the current program state at \texttt{Start}, switch to a shadow copy of the state, execute the shadow fragment and collect observations, and restore the normal execution of the program when we reach the execution point marked with \texttt{End}.

While our new implementation of the refinement technique can deal with more complex cases, e.g., programs with nested branches, it has some limitations, as shown in the snippet below. Since we negate all conditions in the shadow copies, in the example, we end up synthesizing relations that are unsatisfiable for the path that accesses memory in the inner `\texttt{if}' statement's true branch.

\begin{figure}[h]
\centering
\includegraphics[width=0.95\linewidth]{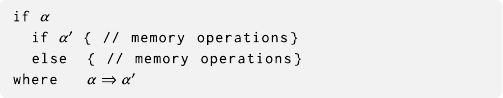}
\vspace{-0.5em}
\Description{Example of a limitation of the refinement technique, in which the negation of the condition of the nested branch makes the relation unsatisfiable.}
\end{figure}

Possible solutions to this problem are to either apply refinement to only the inner `\texttt{if}' statement or implement a heuristics that chooses the right `\texttt{if}' statement (i.e., the one that makes the relation satisfiable) that needs to be refined. For complex scenarios, e.g., with nested branches and functions, similar heuristics could be designed to automatically try different combinations of branches to find optimal refinement settings. We have tried the former solution in our experiments in Sec.~\ref{sec:evaluation}; e.g., to refine observations of Case \#5 when compiled with \opttwo{} enabled. 

The refinement involves two observational models: the base model $\obsmodel{}$ and its refined counterpart $\obsmodel{}'$. We follow the technique of Buiras et al.~\cite{scamv:buiras2021} to optimize the process of computing the observations w.r.t. the two models.
Say $l^{\obsmodel{}}$ and $l^{\obsmodel{}'}$ are, respectively, the observation lists obtained by symbolically executing the program under the base and the refined models. As Fig.~\ref{fig:running:eg:shadow:code} shows $l^{\obsmodel{}} \subseteq l^{\obsmodel{}'}$. Thus, we can implement a projection function $\projname$ that enables obtaining the symbolic observation list of the base model from the observation list of the refined model, i.e., $\projection{l^{\obsmodel{}'}}{l^{\obsmodel{}}}$, without the need of executing symbolic execution twice.

\subsection{Branch misprediction}
To ensure that misprediction will happen in the desired branch, we must train the branch predictor. The standard observation model to check a program's vulnerability to transient execution observes both the \emph{program counter} and the \emph{address of memory operations}. Therefore, each pair of
test input states, i.e., two observational equivalent states, follow the same execution path and satisfy the same path condition $\mathbf{p}$. Based on this insight, to generate a training state, it is sufficient to find satisfying assignments for a different path $\mathbf{p'}$ in the symbolic execution tree, where $\mathbf{p \neq p'}$.

\subsection{Microarchitectural state configuration}
\label{sec:micro-state-config}
The state of microarchitectural features can impact the success of side-channel attacks such as Spectre, as it may affect the speculative execution process. The presence or absence of certain data in the cache, or the prediction made by the branch predictor, can alter the speculative execution path, leading to successfully mounting that attack or missing potential information leakage. As an instance, in our running example,  the speculation of the nested branch can be affected by the presence of $A[idx]$ in the cache before executing the program.
Therefore, tests that only start with an empty cache may not reflect the actual leakage of the processor (see Sec.~\ref{sec:cacheconfig}).
Furthermore, we have primed the branch predictor state by training it according to different training inputs that we generated based on symbolic execution of the test programs.

\subsection{Optimizing programs hardening}\label{subsec:fenceplacement}
Given a program fragment to analyze, the base
and the one with shadow observations represents two observational models: the first one models leakage of a non-speculative processor, and the latter models leakage of a processor that can leak information by speculatively executing all shadow instructions.
Assuming that the fragment is large enough to fill the processor's speculative window, the second model represents the worst-case scenario from a defense point of view, where all potential speculative memory operations must be protected.

On a real processor, the worst-case scenario is not usually possible: the processor may not be able to execute all the speculative instructions. For example, peculiarities of the program may prevent some cache misses, and inter-instruction dependencies may limit the ability of the processor to proceed with speculative execution.

We can generate different observational models by removing some of the shadow observations. We say that
model $M_{1}$ refines model $M_{2}$ if any pair of states that is $\sim_{M_{2}}$, is also $\sim_{M_{1}}$. For example, it is easy to show that if model $M_{2}$ is obtained by removing a shadow observation of model $M_{2}$, then $M_{1}$ refines $M_{2}$.  This forms a lattice of models, where the bottom corresponds to the original program and the top corresponds to the program with all shadow observations. Navigating this lattice can be used to
identify which shadow observation (i.e., speculative instruction) causes the leakage and must be protected.
To guide our optimizing \slh{} hardening, we adopt this lattice structure. 
Intuitively, for a given program and on a specific processor, there is a hardened copy that is produced conservatively and fully protects against transient leakages. On the opposite side, we have a program that is not hardened and is vulnerable to transient leakages. Between these two, we can find many other \emph{partially} hardened programs.  Our technique is to walk---from the fully hardened version to not protected one---on this lattice to find a version of a program which is hardened with a maximally permissive set of fences, i.e., removing any additional hardening would lead to leakage of secret data in some form.

\subsection{Classification aware tests}
We would also like to avoid the introduction of
protections against leakage of variables that are
already public. For example, without knowing the classification of variables for the running example, we should consider the load at line 24 of the \optzero{} compilation potentially insecure, since it may leak the value of $idx$.
However, if $idx$ is public, we should avoid generating experiments where the index differs in $s_{1}$ and $s_{2}$, since this may lead to useless
experiments where the potential different cache footprint depends on public information.
Therefore, we must allow users to possibly configure
variable classification and add the constraint $s_{1} =_{L} s_{2}$ to the relation generated by \scamv{}.
We achieve this by allowing the user to add arbitrary initial observations to the translated program into the BIR language. Since \scamv{} generates only pairs of states that are observationally equivalent, adding
an initial observation for each public variable results in restricting \scamv{} to generate only a pair of states that are low-equivalent.

%%% Local Variables:
%%% mode: latex
%%% TeX-master: "../main"
%%% End:

\section{Implementation}
\label{sec:implementation}

\scamv{}\footnote{\scamv{} is available at: \url{https://github.com/FMSecure/HolBA/tree/dev_scamv_spec}} is developed as a part of HolBA~\cite{trabin} using HOL4's meta language SML. We have extended and modified the \scamv{'s} pipeline implementation to
(1) extend coverage of transient execution vulnerabilities;
(2) analyze binaries produced by common compilers;
(3) tailor countermeasures to specific microarchitectures.

\label{subsec:scamvchanges}

\subsection{\angr{} integration} \label{subsec:angrintegration}
The internal symbolic execution engine of \scamv{} does not scale even to mid-size (more than ten instructions) programs, and the infeasible paths are not pruned in its execution tree.
These render \scamv{} impractical to analyze real-world programs; e.g., Spectre-PHT gadget extracted from OpenSSL in Sec.~\ref{sec:crypto:analysis} consists of a series of conditional statements that cannot be handled by \scamv{'s} symbolic execution. In order to resolve these issues, we have integrated \scamv{} to \angr{}~\cite{angr}---a state-of-the-art binary analysis framework. This required several changes ranging from developing a new interface for communication between \scamv{} and \angr{} to modifying \scamv{'s} pipeline to work with the \angr{} generated symbolic execution tree.

\subsubsection{ \bir{} to \vex{} transpilation}
To outsource symbolic execution to \angr{}, we have implemented a translation from \scamv{'s} intermediate language \bir{} to \vex{}, which is a representation used by the \angr{} internal analysis passes.
On the other hand, the output of \angr{} symbolic execution is the list of observations and the set of path constraints that are expressed in Claripy abstract syntax tree\footnote{Claripy is a Python library for constraint-based symbolic execution.}. Therefore, to transfer back the results of the \angr{} symbolic execution, we also had to implement a translation from Claripy in \angr{} into \bir{}.
Interfacing \scamv{} and \angr{} further required extending \bir{} to support missing \vex{} constructs like \emph{bitstring concatenation}.  

\subsubsection{Handling observations in \angr{}}
In contrast to \bir{}, \vex{} lacks support for observations that we need in our analysis. To compensate for this shortcoming, the translation module from \bir{} to \vex{} replaces \bir{} observations with an \angr{} \emph{system call}, a feature of \angr{} to modify the symbolic state, handle library calls, etc.
We have implemented a specific \angr{} system call handler to process observations. The handler takes as input the state elements we want to observe, like \emph{memory address} or \emph{register number}, and updates the list of observations in the symbolic state for the running path.

\subsubsection{Simulating speculative execution in \angr{}}
\begin{figure}
  \includegraphics[width=0.85\columnwidth]{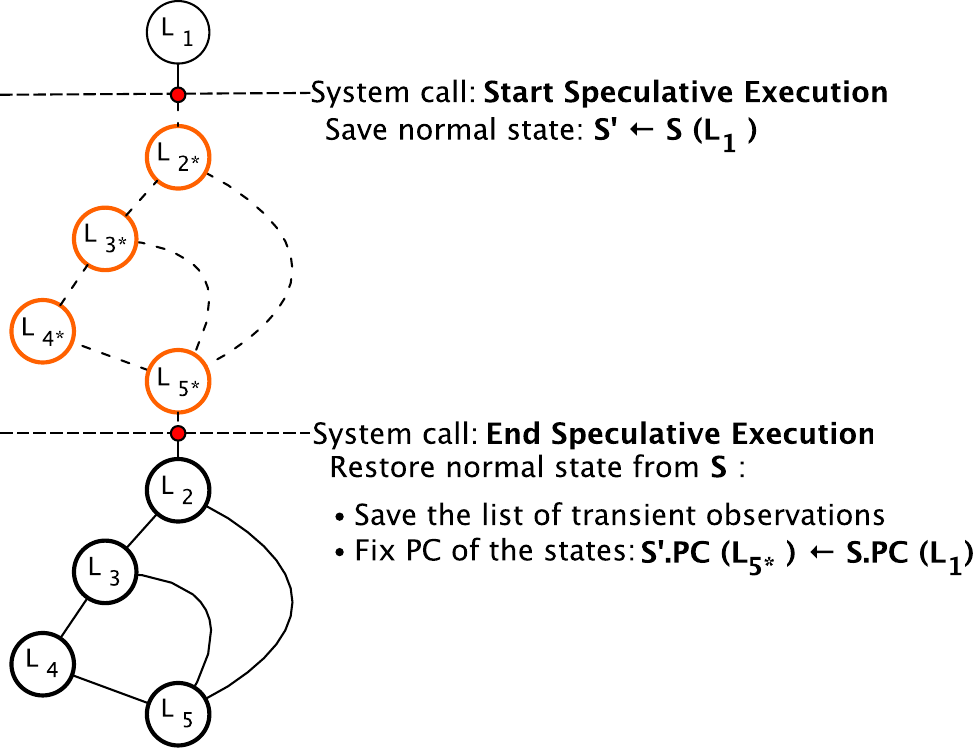}
  \caption{\angr{'s} system calls marking speculative execution of the running example. $S$ denotes the system state.}
  \label{fig:angr-shadow-execution}
  \Description{Abstract graph of the angr symbolic execution, simulating speculative execution through angr system calls.}
\end{figure}

We have also used system calls to simulate speculative execution up to a parameterized depth $d$, reminiscent of the processor speculation depth, in the \angr{} symbolic execution. As shown in Fig.~\ref{fig:angr-shadow-execution}, we use two system calls to mark where the speculative execution begins and ends. The first system call saves the current state using \textit{global plugin} feature of the \angr{} symbolic execution to maintain the program state across multiple execution paths. Then we jump into the code fragment that is supposed to run speculatively.  Having reached the specified depth $d$, we use the second system call to collect the transient observations and the path constraints and then context-switch to the normal execution by restoring the program state prior to the start of speculative execution.

\subsubsection{Concretization of memory accesses}
\angr{} is a static and dynamic symbolic (a.k.a \emph{concolic}) executor based on the Z3 solver. It trades performance for soundness by adopting \textsc{Mayhem}~\cite{mayhem} \emph{partial memory model} to scale to large codebases. Using this model, all symbolic pointers used in memory \emph{store} operations are \emph{concretized} by making a query to Z3. However, the address of \emph{load} operations is conditionally and based on the size of the contiguous interval of possible values treated as symbolic or get concretized. 

Concretization makes symbolic execution more efficient. Yet, to build a generic equivalence relation in the BIR language that can be used to generate multiple test cases by querying an SMT solver, we would need a strategy to generalize from concretized memory locations and perform a remapping from concrete to symbolic values to the corresponding BIR symbolic expressions. For memory addresses, \angr{} keeps a mapping (in the path predicates) from concrete values to the corresponding symbolic expressions, which facilitates the reconstruction of symbolic expressions.

\begin{figure}
\centering
\includegraphics[width=0.9\columnwidth]{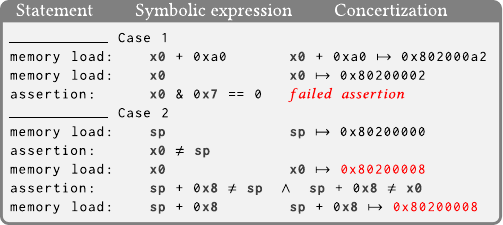}
\caption{\angr{} problem. Case \#1: inconsistency of
concretizations with the following code assertion. Case \#2: the collision of \angr{} concretizations.}
\label{fig:concretization:example1}
\vspace{-1em}
\Description{Two cases of problems encountered in angr. Case 1: concretization of x0 causes a subsequent alignment assertion of x0 to fail. Case 2: two different variables are concretized with two different values, but a subsequent concretization of one of these variables plus an offset collides with the other variable making the path unsatisfiable.}
\end{figure}

Nevertheless, the existing \angr{} \textit{concretization strategy} is insufficient for our analysis. For example, for the load operations, \angr{} queries Z3 for the min and max value of memory addresses under the current path conditions, but our analysis expects one specific valuation of addresses. If the concretization fails, \angr{} marks the memory as \emph{unconstrainted}, which also breaks our analysis. Moreover, since \angr{} uses a \emph{just-in-time} style strategy to concretize symbolic expressions, if the concretized values invalidate an assertion that comes later on a path, \angr{} prunes the path rather than restarting the concretization. Case \#1 in Fig.~\ref{fig:concretization:example1} is an example where concretizations of two memory addresses are consistent with each other, but a following alignment check makes the path unsatisfiable. 
 
Moreover, \angr{} uses a na\"ive strategy to implement concretization, and it always produces a new value for symbolic expressions it encounters. This, however, usually breaks the consistency of values assigned to a specific expression.
Finally, the \angr{} concretization is not collision free, which may cause different symbolic expressions to be mapped to the same value. Case \#2 in Fig.~\ref{fig:concretization:example1} is an example showing how   \texttt{x0} and \texttt{sp} registers are mapped to the same value by the \angr{} concretization.
Given these problems, to ensure the soundness of our approach, we had to develop an efficient concretization strategy that does not suffer from these limitations. 

\textbf{New concretization strategy:}
For a symbolic memory address $a_i$ and a path constraint $\phi_i$ associated with $a_i$, the concretization of $a_i$ is performed by submitting a query $q_{i}$ that is constrained with $\phi_i$ to the SMT solver. The obtained concrete value $s_i$ is assigned to $a_i$ and added as a new constraint to the path. This will ensure that solutions for the subsequent queries will be consistent with those obtained in the previous steps. Additional constraints are also provided to ensure that solutions do not collide with each other.
More formally, if a solution $s_i$ is obtained for a path constraint $\phi_i$ associated with memory access $(a_i, \phi_i)$ in the sequence of queries submitted to the SMT solver, then $s_i$ does not match any previous solution $s_j$ obtained for a path constraint $\phi_j$ associated with memory access $(a_j, \phi_j)$ with $j<i$.

We repeat the concretization each time a memory access is encountered during the symbolic execution. We do not query the solver for memory addresses that have already undergone concretization. Also, a record of all performed concretizations is maintained in order to retrieve old solutions.

If a solution cannot be found for a query, then the SMT solver will be asked to find a solution for all the memory accesses reached up to that point with one single query. If such a solution is found, the concretization of the memory access for the path that fails is updated accordingly and the symbolic execution is restarted from the beginning. To avoid already explored paths, we add the path constraints of the failed path to the initial state. If no solution is found, the concretization of the program is deemed impossible and the path will be pruned from the execution.

\subsection{Optimizing the LLVM's \slh{} hardening}
\label{subsec:fenceplacementimpl}

\paragraph{Selective SLH} Our \slh{} optimization algorithm takes as input a version of the program that is fully hardened. We enumerate the introduced hardening in the program by changing the LLVM pass. Then \scamv{} tries to find hardenings that are not required based on the leakage lattice of Sec.~\ref{subsec:fenceplacement} when the program executes on a specific platform.  
Essentially, our algorithm removes the \slh{} hardening (or poisoning) of loads from top to bottom to keep only those that are essential to stop the leakage. 
Masks introduced by \slh{} are independent of each other, therefore removing fences with their corresponding mask operation does not affect the subsequent fences. Yet, to ensure the soundness of this optimization our algorithm does not remove instructions that perform taint tracking, which is essential for \slh{}.
Algo.~\ref{alg:selective-slh} presents our optimization algorithm. In Algo.~\ref{alg:selective-slh}, ``$\mathit{hasSideChannelLeakage}$'' invokes \scamv{} to test the program under test for the existence of any leakages.

\begin{algorithm}
\caption{\scamv{} Selective SLH}
\label{alg:selective-slh}
\begin{flushleft}
\textbf{Input}: program P hardened using LLVM \slh{}\\
\textbf{Output}: program P' with an optimized number of hardenings \\
\end{flushleft}

\begin{algorithmic}[1]
\PROCEDURE{SelectiveSLH}{$P$}
    \STATE $P' \gets P$ \COMMENT{Initialize $P'$ with the original program}
    \STATE $EnH \gets \mathit{enumerateHardening}(P')$ \COMMENT{List all load hardenings}
    \FOR{$i \gets 1$ to $\mathit{length}(EnH)$}
        \STATE $\mathit{RemoveHrd}(P', EnH[i])$ \COMMENT{Remove the $i$-th hardening}
        \IF{$\mathit{hasSideChannelLeakage}(P')$} 
            \STATE $\mathit{insertHrd}(P', EnH[i])$ \COMMENT{Reinsert the $i$-th hardening}
        \ENDIF
    \ENDFOR
    \RETURN $P'$
\ENDPROCEDURE
\end{algorithmic}
\end{algorithm}

\subsection{Experiment setup} 
\label{para:experiment-setup}
We have run our test cases, each consisting of a program and two inputs, under \textbf{seven} different cache configurations, starting with an empty cache in the first iteration to mimic a cold start. However, for the subsequent iterations, we replicate a more realistic execution environment to account for the effects of cache hits and misses by constructing a cache state based on its content after the first iteration and randomly evicting cache lines.
This is to ensure that speculative execution does not get trapped in other loads in the mispredicted branch and, with a higher probability, can reach interesting loads that leak secret data. 

To ensure the consistency of our results for each cache configuration, we executed each experiment \textbf{ten} times, and we used the same cache state for runs starting from the two inputs.
Unless all these ten executions give the same result, the experiment is classified as inconclusive. In order to resolve such cases, we inspect the cache state after each iteration and we keep the count of valid cache lines, which are populated by the program.
Based on the collected data, a counterexample happens when a cache line was present in 70\% of ten iterations in the cache state of one run, and it never appeared in the cache for the other run.
In case all valid cache lines were present in the cache state of both runs at least once, we mark the experiment as \emph{conclusive} when the total number of cache lines that were present in the cache state of both runs is at least 80\% of the total number of all valid cache lines in both runs. We have chosen this threshold based on our statistical analysis to exclude outliers.

%%% Local Variables:
%%% mode: latex
%%% TeX-master: "../main"
%%% End:

\section{Evaluation}
\label{sec:evaluation}
\renewcommand*{\thefootnote}{\fnsymbol{footnote}}

\begin{table*}[]
\huge
\centering
\resizebox{\textwidth}{!}{%
\begin{tabular}{c|c|c!{\vrule width 2pt}c|c|c|c|c!{\rule{0.5pt}{10pt}}c!{\rule{0.5pt}{10pt}}c|c!{\rule{0.5pt}{10pt}}c!{\rule{0.5pt}{10pt}}c|c!{\rule{0.5pt}{10pt}}c!{\rule{0.5pt}{10pt}}c|c|c|c|c|c!{\rule{0.5pt}{10pt}}c!{\rule{0.5pt}{10pt}}c|c!{\rule{0.5pt}{10pt}}c!{\rule{0.5pt}{10pt}}c|c!{\rule{0.5pt}{10pt}}c!{\rule{0.5pt}{10pt}}c}
\hline
\multirow{2}{*}{\textbf{Case}} &  & \multirow{2}{*}{\textbf{\#Exp}} & \multicolumn{13}{c!{\vrule width 2pt}}{\textbf{Cortex-A53}} & \multicolumn{13}{c}{\textbf{Cortex-A72}} \\ \cline{4-29}
 &  &  & \textbf{\#C} & \textbf{\#I} & \textbf{\#SLH} & \textbf{\#OpSLH} & \multicolumn{3}{c|}{\textbf{ExT}} & \multicolumn{3}{c|}{\textbf{SLHExT}} & \multicolumn{3}{c!{\vrule width 2pt}}{\textbf{OSLHExT}} & \textbf{\#C} & \textbf{\#I} & \textbf{\#SLH} & \textbf{\#OpSLH} & \multicolumn{3}{c|}{\textbf{ExT}} & \multicolumn{3}{c|}{\textbf{SLHExT}} & \multicolumn{3}{c}{\textbf{OSLHExT}} \\ 
 \hline
\multirow{2}{*}{01} & -O0 & 500 & 0 & 0 & 6 & / & 929&928&0.58 & 932&931.57&0.79 & \multicolumn{3}{c!{\vrule width 2pt}}{/} & \cellcolor{red!20}241 & \cellcolor{red!20}259 & \cellcolor{red!20}6 & \cellcolor{red!20}0 & \cellcolor{red!20}864&\cellcolor{red!20}862.29&\cellcolor{red!20}1.60 & \cellcolor{red!20}948&\cellcolor{red!20}946&\cellcolor{red!20}1.29 & \cellcolor{red!20}966&\cellcolor{red!20}963.57&\cellcolor{red!20}1.51 \\
 & -O2 & 500 & 0 & 0 & 4 & / & 688&688&0.0 & 904&903.29&0.49 & \multicolumn{3}{c!{\vrule width 2pt}}{/} & 477 & 4 & 4 & 1 & 686&685.42&0.53 & 962&949.71&6.85 & 962&943.14&14.36 \\
\hline
 \multirow{2}{*}{02} & -O0 & 500 & 0 & 0 & 7 & / & 920&919.43&0.53 & 941&940.43&0.53 & \multicolumn{3}{c!{\vrule width 2pt}}{/} & 0 & 20 & 7 & / & 976&972.43&2.51 & 955&952&1.63 & \multicolumn{3}{c}{/} \\
  & -O2 & 500 & 0 & 0 & 4 & / & 688&688&0.0 & 906&904.86&0.69 & \multicolumn{3}{c!{\vrule width 2pt}}{/} & 477 & 5 & 4 & 1 & 686&685.43&0.53 & 964&945.43&10.63 & 974&948&18.64 \\
 \hline
\multirow{2}{*}{03} & -O0 & 500 & 0 & 0 & 8 & / & 1183&1183&0.0 & 1459&1459&0.0 & \multicolumn{3}{c!{\vrule width 2pt}}{/} & 0 & 0 & 8 & / & 1305&1304.71&0.49 & 1565&1564.57&0.53 & \multicolumn{3}{c}{/} \\
 & -O2 & 500 & 0 & 0 & 4 & / & 917&916.57&0.53 & 1036&1035.86&0.38 & \multicolumn{3}{c!{\vrule width 2pt}}{/} & \cellcolor{blue!20}28 & \cellcolor{blue!20}51 & \cellcolor{blue!20}4 & \cellcolor{blue!20}1 & \cellcolor{blue!20}935&\cellcolor{blue!20}934.14&\cellcolor{blue!20}0.38 & \cellcolor{blue!20}1063&\cellcolor{blue!20}1059&\cellcolor{blue!20}3.87 & \cellcolor{blue!20}938&\cellcolor{blue!20}937.29&\cellcolor{blue!20}0.76 \\
\hline
\multirow{2}{*}{04} & -O0 & 500 & 0 & 0 & 6 & / & 929&928.29&0.49 & 932&930.86&0.69 & \multicolumn{3}{c!{\vrule width 2pt}}{/} & \cellcolor{red!20}37 & \cellcolor{red!20}52 & \cellcolor{red!20}6 & \cellcolor{red!20}0 & \cellcolor{red!20}866&\cellcolor{red!20}862&\cellcolor{red!20}2.24 & \cellcolor{red!20}1063&\cellcolor{red!20}1058.14&\cellcolor{red!20}4.45 & \cellcolor{red!20}965&\cellcolor{red!20}962.71&\cellcolor{red!20}1.70 \\
 & -O2 & 156 & 0 & 0 & 4 & / & 793&792.43&0.53 & 907&906.43&0.79 & \multicolumn{3}{c!{\vrule width 2pt}}{/} & 122 & 20 & 4 & 1 & 800&799.86&0.38 & 863&860.57&1.72 & 977&941.86\footnotemark[2]&19.19 \\
\hline
\multirow{2}{*}{05} & -O0 & 500*3 & 0 & 0 & 9 & / & 1272&1271.71&0.49 & 1437&1436.43&0.53 & \multicolumn{3}{c!{\vrule width 2pt}}{/} & 0 & 0 & 9 & / & 1150&1147&2.08 & 1548&1548&0.0 & \multicolumn{3}{c}{/} \\
 & -O2 & 500*2 & 0 & 0 & 4 & / & 794&793.14&0.38 & 1043&1042.43&0.53 & \multicolumn{3}{c!{\vrule width 2pt}}{/} & \cellcolor{red!20}349 & \cellcolor{red!20}245 & \cellcolor{red!20}4 & \cellcolor{red!20}0 & \cellcolor{red!20}820&\cellcolor{red!20}817&\cellcolor{red!20}1.91 & \cellcolor{red!20}992&\cellcolor{red!20}989.86&\cellcolor{red!20}1.07 & \cellcolor{red!20}850&\cellcolor{red!20}848.57&\cellcolor{red!20}1.13 \\
\hline
\multirow{2}{*}{06} & -O0 & 500 & 0 & 0 & 7 & / & 932&931&0.58 & 1055&1053.71&0.76 & \multicolumn{3}{c!{\vrule width 2pt}}{/} & 0 & 0 & 7 & / & 936&934.86&0.90 & 1174&1173.29&0.49 & \multicolumn{3}{c}{/} \\
 & -O2 & 500 & 0 & 0 & 4 & / & 793&793&0.0 & 1048&1047.14&0.38 & \multicolumn{3}{c!{\vrule width 2pt}}{/} & \cellcolor{blue!20}482 & \cellcolor{blue!20}0 & \cellcolor{blue!20}4 & \cellcolor{blue!20}1 & \cellcolor{blue!20}819&\cellcolor{blue!20}817.14&\cellcolor{blue!20}1.35 & \cellcolor{blue!20}985&\cellcolor{blue!20}982.85&\cellcolor{blue!20}1.21 & \cellcolor{blue!20}827&\cellcolor{blue!20}825.43&\cellcolor{blue!20}1.13 \\
\hline
\multirow{2}{*}{07} & -O0 & 500*2 & 0 & 0 & 9 & / & 1308&1308&0.0 & 1187&1186.43&0.53 & \multicolumn{3}{c!{\vrule width 2pt}}{/} & \cellcolor{red!20}330 & \cellcolor{red!20}25 & \cellcolor{red!20}9 & \cellcolor{red!20}0 & \cellcolor{red!20}1336&\cellcolor{red!20}1335&\cellcolor{red!20}0.82 & \cellcolor{red!20}1172&\cellcolor{red!20}1170.71&\cellcolor{red!20}0.95 & \cellcolor{red!20}1339&\cellcolor{red!20}1337.14&\cellcolor{red!20}1.21 \\
 & -O2 & 500*2 & 0 & 0 & 5 & / & 784&783.71&0.49 & 1163&1162.43&0.53 & \multicolumn{3}{c!{\vrule width 2pt}}{/} & 498 & 0 & 5 & 1 & 807&805.71&1.11 & 1037&1036.57&0.53 & 1173&1172.14\footnotemark[2]&0.69 \\
\hline
\multirow{2}{*}{08} & -O0 & 1000 & 0 & 0 & 7 & / & 912&911.29&0.49 & 963&961.71&0.76 & \multicolumn{3}{c!{\vrule width 2pt}}{/} & 0 & 43 & 7 & / & 871&867.29&1.98 & 1008&1007&0.58 & \multicolumn{3}{c}{/} \\
 & -O2 & / & / & / & / & / & \multicolumn{3}{c|}{/} & \multicolumn{3}{c|}{/} & \multicolumn{3}{c!{\vrule width 2pt}}{/} & / & / & / & / & \multicolumn{3}{c|}{/} & \multicolumn{3}{c|}{/} & \multicolumn{3}{c}{/} \\
\hline
\multirow{2}{*}{09} & -O0 & / & / & / & / & / & \multicolumn{3}{c|}{/} & \multicolumn{3}{c|}{/} & \multicolumn{3}{c!{\vrule width 2pt}}{/} & / & / & / & / & \multicolumn{3}{c|}{/} & \multicolumn{3}{c|}{/} & \multicolumn{3}{c}{/} \\
& -O2 & / & / & / & / & / & \multicolumn{3}{c|}{/} & \multicolumn{3}{c|}{/} & \multicolumn{3}{c!{\vrule width 2pt}}{/} & / & / & / & / & \multicolumn{3}{c|}{/} & \multicolumn{3}{c|}{/} & \multicolumn{3}{c}{/} \\
\hline
\multirow{2}{*}{09v2} & -O0 & / & / & / & / & / & \multicolumn{3}{c|}{/} & \multicolumn{3}{c|}{/} & \multicolumn{3}{c!{\vrule width 2pt}}{/} & / & / & / & / & \multicolumn{3}{c|}{/} & \multicolumn{3}{c|}{/} & \multicolumn{3}{c}{/} \\
& -O2 & / & / & / & / & / & \multicolumn{3}{c|}{/} & \multicolumn{3}{c|}{/} & \multicolumn{3}{c!{\vrule width 2pt}}{/} & / & / & / & / & \multicolumn{3}{c|}{/} & \multicolumn{3}{c|}{/} & \multicolumn{3}{c}{/} \\
\hline
\multirow{2}{*}{10} & -O0 & 500*3 & 0 & 0 & 7 & / & 1173&1173&0.0 & 1236&1236&0.0 & \multicolumn{3}{c!{\vrule width 2pt}}{/} & 0 & 1 & 7 & / & 1065&1063.57&0.79 & 1185&1184.29&0.49 & \multicolumn{3}{c}{/} \\
 & -O2 & 500*3 & 0 & 0 & 4 & / & 601&601&0.0 & 965&964.86&0.38 & \multicolumn{3}{c!{\vrule width 2pt}}{/} & 0 & 155 & 4 & / & 696&696&0.0 & 1023&1014.86&7.03 & \multicolumn{3}{c}{/} \\
\hline
\multirow{2}{*}{11gcc} & -O0 & 500 & 0 & 0 & 18 & / & 2046&2045.57&0.53 & 2733&2732.29&0.49 & \multicolumn{3}{c!{\vrule width 2pt}}{/} & 0 & 0 & 18 & / & 2162&2160.86&0.69 & 2688&2687.29&0.95 & \multicolumn{3}{c}{/} \\
 & -O2 & 500 & 0 & 0 & 4 & / & 789&788.43&0.53 & 1051&1050.86&0.38 & \multicolumn{3}{c!{\vrule width 2pt}}{/} & \cellcolor{blue!20}479 & \cellcolor{blue!20}1 & \cellcolor{blue!20}4 & \cellcolor{blue!20}1 & \cellcolor{blue!20}819&\cellcolor{blue!20}816.86&\cellcolor{blue!20}1.57 & \cellcolor{blue!20}1045&\cellcolor{blue!20}1044.71&\cellcolor{blue!20}0.49 & \cellcolor{blue!20}865&\cellcolor{blue!20}863.14&\cellcolor{blue!20}1.34 \\
\hline
\multirow{2}{*}{11ker} & -O0 & 500*2 & 0 & 0 & 17 & / & 2183&2182.14&0.38 & 2760&2760&0.0 & \multicolumn{3}{c!{\vrule width 2pt}}{/} & 0 & 0 & 17 & / & 2162&2160.71&0.95 & 2702&2700.71&0.76 & \multicolumn{3}{c}{/} \\
 & -O2 & 500 & 0 & 0 & 4 & / & 688&688&0.0 & 906&905.43&0.79 & \multicolumn{3}{c!{\vrule width 2pt}}{/} & 476 & 5 & 4 & 1 & 686&685.86&0.38 & 964&944.57&15.11 & 966&948.43&12.99 \\
\hline
\multirow{2}{*}{11sub} & -O0 & 500 & 0 & 0 & 22 & / & 2076&2075.14&0.38 & 2720&2719.71&0.49 & \multicolumn{3}{c!{\vrule width 2pt}}{/} & 0 & 0 & 22 & / & 2060&2059.71&0.49 & 2565&2563.86&1.21 & \multicolumn{3}{c}{/} \\
 & -O2 & 500 & 0 & 0 & 4 & / & 688&688&0.0 & 906&904.71&0.95 & \multicolumn{3}{c!{\vrule width 2pt}}{/} & 476 & 2 & 4 & 1 & 686&685.43&0.53 & 962&945.14&12.43 & 966&945.57&10.10 \\
\hline
\multirow{2}{*}{12} & -O0 & 500 & 0 & 0 & 8 & / & 937&936.43&0.53 & 1176&1176&0.0 & \multicolumn{3}{c!{\vrule width 2pt}}{/} & 0 & 0 & 8 & / & 907&905.86&0.90 & 1142&1139.86&2.27 & \multicolumn{3}{c}{/} \\
 & -O2 & 500 & 0 & 0 & 4 & / & 703&703&0.0 & 834&834&0.0 & \multicolumn{3}{c!{\vrule width 2pt}}{/} & 345 & 80 & 4 & 1 & 812&812&0.0 & 863&861.29&1.38 & 938&937\footnotemark[2]&1.15 \\
\hline
\multirow{2}{*}{13} & -O0 & 500 & 0 & 0 & 8 & / & 1012&1012&0.0 & 1434&1434&0.0 & \multicolumn{3}{c!{\vrule width 2pt}}{/} & 0 & 0 & 8 & / & 1167&1165.57&1.13 & 1325&1324.71&0.49 & \multicolumn{3}{c}{/} \\
 & -O2 & 500 & 0 & 0 & 4 & / & 688&688&0.0 & 905&904.29&0.76 & \multicolumn{3}{c!{\vrule width 2pt}}{/} & 477 & 5 & 4 & 1 & 686&685.28&0.49 & 958&945.85&11.58 & 966&939.42&14.46 \\
\hline
\multirow{2}{*}{14} & -O0 & 500 & 0 & 0 & 6 & / & 929&928.57&0.53 & 931&930.86&0.38 & \multicolumn{3}{c!{\vrule width 2pt}}{/} & \cellcolor{red!20}1 & \cellcolor{red!20}5 & \cellcolor{red!20}6 & \cellcolor{red!20}0 & \cellcolor{red!20}866&\cellcolor{red!20}861.57&\cellcolor{red!20}2.64 & \cellcolor{red!20}1057&\cellcolor{red!20}1053.86&\cellcolor{red!20}2.12 & \cellcolor{red!20}964&\cellcolor{red!20}961.86&\cellcolor{red!20}1.95 \\
 & -O2 & 500 & 0 & 0 & 4 & / & 793&792.43&0.53 & 906&905.29&0.49 & \multicolumn{3}{c!{\vrule width 2pt}}{/} & 500 & 0 & 4 & 1 & 800&800&0.0 & 862&860.57&0.98 & 960&947.86\footnotemark[2]&14.15 \\
\hline
\multirow{2}{*}{14v2} & -O0 & 500 & 0 & 0 & 8 & / & 931&930.71&0.49 & 1186&1185.86&0.38 & \multicolumn{3}{c!{\vrule width 2pt}}{/} & 0 & 0 & 8 & / & 1039&1027.71&6.80 & 1312&1310.86&0.69 & \multicolumn{3}{c}{/} \\
 & -O2 & 500 & 0 & 0 & 4 & / & 792&791.43&0.53 & 1047&1047&0.0 & \multicolumn{3}{c!{\vrule width 2pt}}{/} & 0 & 0 & 4 & / & 819&816.71&1.98 & 1045&1043&1.41 & \multicolumn{3}{c}{/} \\
\hline
\multirow{2}{*}{SiSCloack} & -O0 & 500 & 0 & 0 & 3 & / & 1024&1024&0.0 & 1145&1144.43&0.53 & \multicolumn{3}{c!{\vrule width 2pt}}{/} & 422 & 0 & \textcolor{red}{3} & / & \multicolumn{3}{c|}{/} & \multicolumn{3}{c|}{/} & \multicolumn{3}{c}{/} \\
 & -O2 & 500 & 488 & 0 & 1 & / & 579&579&0.0 & 798&798&0.0 & \multicolumn{3}{c!{\vrule width 2pt}}{/} & 488 & 0 & \textcolor{red}{1} & / & \multicolumn{3}{c|}{/} & \multicolumn{3}{c|}{/} & \multicolumn{3}{c}{/} \\
\hline
\end{tabular}%
}
\caption{ Analysis of collected microbenchmarks~\cite{kocher:sepcterv1-benchmarks,guarnieri:spectector,ctfoundations:pldi:20}. 
Abbreviations in the table: Exp: number of experiments (for some cases, multiplication shows how many times the experiment is repeated with different ways of training the branch predictor), C: Counterexamples, I: Inconclusive cases, SLH: \slh{} inserted hardenings, OpSLH: retained hardening by \scamv{}, (O)SLHExT: (optimized) (SLH) Execution time in CPU cycle. The execution time is represented by a tuple of three values corresponding to the maximum value, the average and the standard deviation, separated by ''|''. Highlighted rows are discussed in Section ~\ref{sec:security:profromance}.}
\label{tab:results:benchmarks}
\vspace{-1em}
\end{table*}

To show the effectiveness of \scamv{} in optimally protecting programs against Spectre-PHT, we have conducted several experiments on two widely used ARMv8 boards, Raspberry Pi 3 and 4.
First, we have evaluated \scamv{} on a suit of benchmarks  that are used by Kocher~\cite{kocher:sepcterv1-benchmarks} and others~\cite{guarnieri:spectector,ctfoundations:pldi:20,Mosier2022} to analyze and mitigate transient execution attacks~\cite{ctfoundations:pldi:20,clou,guarnieri:spectector}. Second,  we used \scamv{} to analyze the vulnerability of the OpenSSL library on AArch64 processors and optimally harden it to stop found leakages. For this experiment, we have only analyzed fragments of OpenSSL that are reported to be vulnerable to transient execution attacks in related literature~\cite{lcms:mosier:2022}.
Our experiments run as bare-metal code, i.e., no operating system or other background processes exist. Therefore, our experiments represent the worst-case scenario where no defense is in place and the attacker can inspect the cache state after execution directly using available hardware instructions.

\paragraph{Evaluation boards} Raspberry Pi 3 and 4 use Cortex-A53 and Cortex-A72 processors, resp. Cortex-A53 is an 8-stage pipelined processor with a 2-way superscalar and in-order execution pipeline. Similarly,  Cortex-A72  is a 15-stage pipelined ARMv8 core with a 3-way superscalar and out-of-order execution pipeline.  Both processors support speculative execution based on control flow prediction. However, while ARM Ltd. declared Cortex-A72's vulnerability to Spectre-PHT, it was recently that the vulnerability of Cortex-A53  to transient execution attacks was proved~\cite{scamv}.
\scamv{} uses a special \emph{module} in TrustZone to run experiments. The module sets up memory types (e.g., the cacheability of memory regions), configures the cache's initial state, and probes the cache state after the execution of programs. In a real attack scenario, an attacker can use \emph{performance monitor counter} or PMC for timing analysis.
We also used the PMC to evaluate the performance of our \slh{} optimization.

\footnotetext[2]{
Table~\ref{tab:results:benchmarks} presents some counterintuitive results, where the optimized code exhibits worse performance than the non-optimized one. Further experiments showed that these results are caused by changes in alignment. In fact, replacing unnecessary hardening with NOP instructions results in performance comparable to the standard \slh{}.\label{footnote:code-alignment}}
\renewcommand*{\thefootnote}{\arabic{footnote}}
\setcounter{footnote}{4}

\subsection{Analyzing Spectre microbenchmarks}
We have successfully analyzed 17 variants of Spectre and detected leakage in those that are vulnerable to Spectre-PHT on our evaluation processors. The only exceptions are cases \#9 and its variation \#9v2, which we could not analyze due to the limitation of our approach (Sec.~\ref{sec:limitations} elaborates on this).
Please note that \emph{specification-based} testing and the use of \emph{observation refinement} in \scamv{} helped us to reduce the number of required test cases in our experiments.

We used \clang{} with compiler optimizations \optzero{} and \opttwo{} to produce the binary of microbenchmarks. We repeated the same experiments on both RPi boards that we have to check their vulnerability on these platforms and to protect them optimally. We identified several cases where compilers unnecessarily hardened programs. This provides opportunities for optimization without sacrificing security. Table~\ref{tab:results:benchmarks} summarizes the results of our experiments for the benchmark programs.
Our evaluation is done under the assumption that the code is executed standalone and no other code or vector of attacks is allowed. Note that the microbenchmarks represent different versions of Spectre-PHT at the source code level. However, when compiling with optimization \opttwo{} enabled, sometimes the same binaries are produced for different cases, e.g. Case \#1 and Case \#13.

\paragraph{Timing analysis} We measured the time by reading the \emph{CPU cycle register} of the PMC. To better evaluate the performance, we used inputs that cause programs to take the longest path possible and to go through as much hardening as possible. To minimize the effect of hardware internal noises that cause variations in CPU cycles,  we run the program under test 50000 times and compute the mean value of all iterations. We repeated each measurement seven times and computed the average, the standard deviation and the maximum value of all the executions.

\subsubsection{Cortex-A53 experiments}
None of the {17} cases from Kocher benchmarks resulted in leakage on Cortext-A53, i.e., no need for any protection.
This may be attributed to the absence of register renaming and the short CPU pipeline, which prevents using the result of a speculated load instruction in subsequent operations. The only case that induces leakage on Cortex-A53 is the last case in Table~\ref{tab:results:benchmarks}, which is a variant of SiSCloak presented in~\cite{scamv:buiras2021}. Compared to Spectre-PHT, in SiSCloak only the load which leaks the secret (read from array \texttt{B}) is protected by the bound check and reading from the public array (\texttt{A}) is moved before the \texttt{if} statement:

\begin{figure}[h]
\centering
\includegraphics[width=0.9\columnwidth]{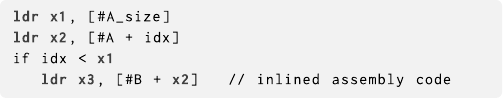}
\Description{Example of SiSCloak vulnerability, where the first load after a conditional branch is vulnerable to Spectre-PHT attack.}
\end{figure}

\slh{} fails to protect against SiSCloak on Cortex-A72 as SiSCloak's implementation relies on inlined assembly that is not considered in the \slh{} protection model. Thus, we do not proceed with further performance analysis of SiSCloak. On Cortex-A53, \slh{} could prevent the leakage, but this was solely due to the added tracking instructions that fill up the processor's short speculation pipeline, causing the leaky load to not execute in speculation.

\subsubsection{Cortex-A72 experiments}
ARM officially confirmed that Cortex-A72 processors are vulnerable to Spectre attacks. However, our experiments highlight interesting findings, which we summarize here.
While for most benchmarks compiled with optimizations \opttwo{} enabled, \scamv{} identified several counterexamples, it only found a few vulnerable cases when \optzero{} was used. This suggests that additional operations, including operations on the stack, which are introduced when the compiler optimizations are disabled, may invalidate the transient leakages.
Also, \scamv{} did not detect any leak for two benchmarks, namely Case \#10 and Case \#14, when compiled with \clang{} using both \optzero{} and \opttwo{}. \scamv{} successfully generated well-formed inputs to exploit transient execution. However, no counterexamples were identified on hardware, even though both cases are considered insecure in the literature~\cite{guarnieri:spectector}. This observation indicates that the properties of the underlying hardware can significantly influence the leakage potential of the code. This is because different architectures may execute the same high-level code using varying machine instructions and ordering, thereby affecting the programs' security.

The other interesting case is Case \#6. The snippet below shows (left)  the unoptimized and unmodified output of \clang{}.
\begin{figure}[h]
\centering
\includegraphics[width=0.9\columnwidth]{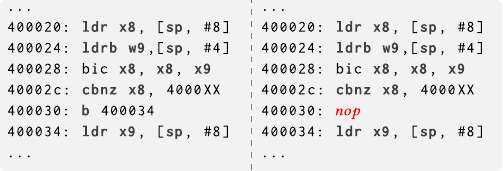}
\Description{Two versions of a snippet of Case 6 in ARM assembly code. The left version illustrates the original code, while the right version shows the same instructions with the replacement of the indirect jump using a NOP instruction.}
\end{figure}
For this version, we could not identify any leaks. However, replacing the jump instruction at address \texttt{400030}, with a \texttt{nop} makes this program vulnerable to Spectre-PHT and \scamv{} identified a few counterexamples. The replaced jump instruction does not change the program control flow and just moves the control to the next instruction in the program order. We conjecture that such jump instructions cause the processor to flush the instruction pipeline.

\subsubsection{Effect of compiler optimizations}
Except Case \#1, \#4, \#7, and \#14, all other cases compiled with \optzero{} show no leakage. When a program is compiled with compiler optimizations disabled (i.e., using \optzero{}), the produced assembly includes several unnecessary memory operations. For example, function arguments, such as the value of the index in the running example, are stored (resp. loaded) on (from) the stack. These additional memory operations can affect speculative execution by filling up the pipeline, causing leakage-inducing operations to not execute in speculation.

\subsubsection{Cache configuration effect} \label{sec:cacheconfig}
As discussed in Sec.~\ref{sec:micro-state-config}, The state of microarchitectural features, like the data cache, can impact the success of
Spectre attacks. We found evidence of this in our microbenchmarks. As an example, consider Case \#1 which leaks data through a memory access like $B[A[idx]]$. When Case \#1 is compiled with \optzero{}, all 241 found counterexamples are achievable only when $A[idx]$ is in the cache. If $A[idx]$ does not hit the cache, fetching $A[idx]$ from the main memory causes a delay, which prevents the execution of $B[A[idx]]$ in speculation.

\subsubsection{Security and performance}\label{sec:security:profromance}
Cases without counterexamples do not need any protection; thus, no hardening optimization was required.
Instead, for those cases that \scamv{} identified counterexamples, we synthesized a minimal set of required hardening needed to protect against data leaks. In particular, we found an improvement in performance in Case \textcolor{blue!65}{\#3}, \textcolor{blue!65}{\#5}, \textcolor{blue!65}{\#6}, \textcolor{blue!65}{\#11gcc} when \slh{} hardening is optimized. Furthermore, we evaluated the security of AArch64 \slh{} and our \slh{} optimization by re-executing counterexamples found for the unprotected program to ensure that the leakage has been mitigated.

Our experiments highlighted a few cases where our optimization pass removes all protections: Case \textcolor{red!65}{\#1}, \textcolor{red!65}{\#4}, \textcolor{red!65}{\#5}, \textcolor{red!65}{\#7}, and \textcolor{red!65}{\#14}. Further analysis revealed that the \slh{} hardening changes the code \textit{alignment} in memory that affects speculative leakage. Code alignment may influence programs' behavior in several ways, such as branch prediction accuracy, memory access patterns, and instruction decoding and dispatching.  For example, if the branch instruction crosses cache line boundaries, this might affect branch prediction accuracy.
Changing memory access patterns could affect, e.g., data prefetching, thus impacting the cache hits rate. 
Also, code alignment may increase the latency of instruction decoding and dispatching, potentially impacting speculative execution.
Notice that code alignment w.r.t. cache lines is preserved by the majority of implementations of ASLR, since randomization is usually done at page-level granularity.

\subsubsection{Analysis limitations} \label{sec:limitations} Our approach has some limitations:
\paragraph{Branch predictor training} To mount Spectre-PHT, we need at least one conditional branch in the program. In this way, one path is necessary to generate an input to train the branch predictor and the opposite one will be used to generate inputs to exploit the miss-prediction. In Case \#9$\_1$, however, we could get only one path from \angr{} due to the constant propagation that leads to pruning of the path. Similarly, \scamv{} could not test Case \#9$\_2$ because of path pruning. The problem arises from utilizing a pointer in the branch condition, which is retrieved from the stack point through a memory load. As we constrain every memory access to be within a specific memory region, the condition will never become false, and therefore, the path is discarded. Case \#8 compiled with \opttwo{} is also excluded from the analysis as it lacks a conditional branch and is therefore not vulnerable to Spectre-PHT. 
\begin{figure}[h]
\centering
\includegraphics[width=0.9\columnwidth]{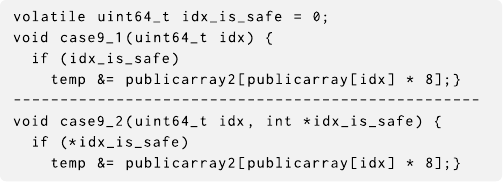}
\Description{Two cases that Scam-V cannot analyze due to the limitations of angr symbolic execution. In Case 9 version 1, the branch condition is a variable assigned to zero, whereas in Case 9 version 2, the branch condition is a pointer.}
\end{figure}

\paragraph{Misspeculation and observation refinement} The other limitation of \scamv{} is associated with applying the observation refinement, as discussed in Sec.~\ref{sec:method}. By default, the misspeculation is expected to trigger at the first conditional branch and continue to all subsequently encountered conditional branches. Thus, our refinement approach negates all branch conditions to make observable memory operations that can be executed in misspeculation. However, this does not always work (e.g., Case \#11$^*$ and \#13), as some attacks do not follow this pattern. To overcome this limitation and ensure that \scamv{} can analyze cases in which potential memory accesses are not always in the opposite branch, we had to manually decide the condition of which \texttt{if} statements must be negated.

\paragraph{Loops in symbolic execution} Handling \emph{infinite loops} is challenging in symbolic execution. In our experiments, only Case \#5 contains an infinite loop, which we handle by performing a one-time unrolling and back-edge cutting of the loop at the LLVM level. For this specific case, one iteration of the loop body was enough to detect the leakage. For all other cases which contain a \emph{finite loop}, we have performed \emph{loop unrolling}.
The same problem can also be addressed by introducing a precondition to constrain variables involved in the loop condition to a specific range.

\subsubsection{More details on Case \#1 and \#10}
\label{sec:eval-specific-cases}

\begin{table*}[]
\huge
\centering
\resizebox{\textwidth}{!}{%
\begin{tabular}{c|c|c!{\vrule width 2pt}c|c|c|c|c!{\rule{0.5pt}{10pt}}c!{\rule{0.5pt}{10pt}}c|c!{\rule{0.5pt}{10pt}}c!{\rule{0.5pt}{10pt}}c|c!{\rule{0.5pt}{10pt}}c!{\rule{0.5pt}{10pt}}c|c|c|c|c|c|c!{\rule{0.5pt}{10pt}}c!{\rule{0.5pt}{10pt}}c|c!{\rule{0.5pt}{10pt}}c!{\rule{0.5pt}{10pt}}c|c!{\rule{0.5pt}{10pt}}c!{\rule{0.5pt}{10pt}}c}
\hline
\multirow{2}{*}{\textbf{Ex.}} &  & \multirow{2}{*}{\textbf{\#Exp}} & \multicolumn{13}{c!{\vrule width 2pt}}{\textbf{Cortex-A53}} & \multicolumn{14}{c}{\textbf{Cortex-A72}} \\ \cline{4-16} \cline{17-30}
 &  &  & \textbf{\#C} & \textbf{\#I} & \textbf{\#H} & \textbf{\#OpH} & \multicolumn{3}{c|}{\textbf{ExT}} & \multicolumn{3}{c|}{\textbf{HExT}} & \multicolumn{3}{c!{\vrule width 2pt}}{\textbf{OpHExT}} & \textbf{\#C} & \textbf{\#I} & \textbf{TypeH} & \textbf{\#H} & \textbf{\#OpH} & \multicolumn{3}{c|}{\textbf{ExT}} & \multicolumn{3}{c|}{\textbf{HExT}} & \multicolumn{3}{c}{\textbf{OpHExT}} \\ 
 \hline
\multirow{2}{*}{01} & -O0 & 500 & 0 & 0 & 6 & / & 928&927.57&0.53
 & 932&931&0.82 & \multicolumn{3}{c!{\vrule width 2pt}}{/} & 464 & 34 &  & 6 & 1 & 863&860.29&1.80 & 948&946.71&1.38 & 1042&1024.14&14.71 \\ 
 & \multirow{2}{*}{-O2} & \multirow{2}{*}{500} & \multirow{2}{*}{0} & \multirow{2}{*}{0} & \multirow{2}{*}{4} & \multirow{2}{*}{/} & \multirow{2}{*}{688}&\multirow{2}{*}{688}&\multirow{2}{*}{0.0} & \multirow{2}{*}{908}&\multirow{2}{*}{906.14}&\multirow{2}{*}{1.07} & \multicolumn{3}{c!{\vrule width 2pt}}{\multirow{2}{*}{/}} & \multirow{2}{*}{500} & \multirow{2}{*}{0} & \cellcolor{blue!10}\textcolor{red}{aSLH} & \cellcolor{blue!10}4 & \cellcolor{blue!10}1 & \cellcolor{blue!10} & \cellcolor{blue!10} & \cellcolor{blue!10} & \cellcolor{blue!10}866&\cellcolor{blue!10}861.71&\cellcolor{blue!10}2.29 & \cellcolor{blue!10}944&\cellcolor{blue!10}937.43&\cellcolor{blue!10}6.45 \\ 
 &  &  &  &  &  &  &  &  &  &  &  &  & \multicolumn{3}{c!{\vrule width 2pt}}{} &  &  & \cellcolor{gray!20}\textcolor{red}{FI} & \cellcolor{gray!20}1 & \cellcolor{gray!20}/ & \cellcolor{gray!20}\multirow{-2}{*}{686}&\cellcolor{gray!20}\multirow{-2}{*}{685.71}&\cellcolor{gray!20}\multirow{-2}{*}{0.49} & \cellcolor{gray!20}999&\cellcolor{gray!20}998.29&\cellcolor{gray!20}0.49 & \multicolumn{3}{c}{\cellcolor{gray!20}/} \\ 
\hline
\multirow{2}{*}{10} & -O0 & 500*3 & 0 & 0 & 7 & / & 936&935.29&0.76 & 1331&1331&0.0 & \multicolumn{3}{c!{\vrule width 2pt}}{/} & 602 & 0 &  & 7 & 1 & 915&912.86&1.77 & 1334&1332.29&1.98 & 1091&1089.43&1.13 \\ 
 & \multirow{2}{*}{-O2} & \multirow{2}{*}{500*3} & \multirow{2}{*}{0} & \multirow{2}{*}{0} & \multirow{2}{*}{4} & \multirow{2}{*}{/} & \multirow{2}{*}{699}&\multirow{2}{*}{698.57}&\multirow{2}{*}{0.53} & \multirow{2}{*}{836}&\multirow{2}{*}{835.86}&\multirow{2}{*}{0.38} & \multicolumn{3}{c!{\vrule width 2pt}}{\multirow{2}{*}{/}} & \multirow{2}{*}{620} & \multirow{2}{*}{216} & \cellcolor{blue!10}\textcolor{red}{aSLH} & \cellcolor{blue!10}4 & \cellcolor{blue!10}1 & \cellcolor{blue!10} & \cellcolor{blue!10} & \cellcolor{blue!10} & \cellcolor{blue!10}1031&\cellcolor{blue!10}1029.57&\cellcolor{blue!10}0.98 & \cellcolor{blue!10}803&\cellcolor{blue!10}802.43&\cellcolor{blue!10}0.53 \\ 
&  &  &  &  &  &  &  &  &  &  &  &  & \multicolumn{3}{c!{\vrule width 2pt}}{} &  &  & \cellcolor{gray!20}\textcolor{red}{FI} & \cellcolor{gray!20}1 & \cellcolor{gray!20}/ & \cellcolor{gray!20}\multirow{-2}{*}{733}&\cellcolor{gray!20}\multirow{-2}{*}{731.29}&\cellcolor{gray!20}\multirow{-2}{*}{1.11} & \cellcolor{gray!20}1049&\cellcolor{gray!20}1046.43&\cellcolor{gray!20}1.51 & \multicolumn{3}{c}{\cellcolor{gray!20}/} \\
\hline
\hline
\multirow{2}{*}{SSL\_get\_shared\_sigalgs} & -O0 & 105*5 & 0 & 0 & 35 & / & 2419&2418.43&0.53 & 3686&3685.43&0.53 & \multicolumn{3}{c!{\vrule width 2pt}}{/} & 0 & 0 &  & 35 & / & 2362&2361&1.41 & 3390&3382.14&4.91 & \multicolumn{3}{c}{/} \\ 
 & -O2 & 85*6 & 0 & 0 & 9 & / & 1190&1190&0.0 & 934&933.29&0.49 & \multicolumn{3}{c!{\vrule width 2pt}}{/} & 78 & 216 & \cellcolor{blue!10}\textcolor{red}{aSLH} & \cellcolor{blue!10}9 & \cellcolor{blue!10}1 & \cellcolor{blue!10}1020&\cellcolor{blue!10}1018.57&\cellcolor{blue!10}0.98 & \cellcolor{blue!10}1094&\cellcolor{blue!10}1093.29&\cellcolor{blue!10}0.49 & \cellcolor{blue!10}1090&\cellcolor{blue!10}1089.43&\cellcolor{blue!10}0.53 \\ 
\hline
\end{tabular}%
}
\caption{Analysis of specific cases for a \textit{private} array index (i.e., \texttt{idx}) and Spectre-PHT gadgets from OpenSSL. 
%See Tab.~\ref{tab:results:benchmarks}. 
\texttt{TypeH} column indicates the type of applied hardening: none (default LLVM \slh{}), aSLH (poisoning on memory addresses), FI (\dsb{}$+$\isb{}).}
\label{tab:further-results}
\vspace{-1em}
\end{table*}

We have investigated Case \#1 (the prime Spectre-PHT example) and Case \#10 (our running example in the paper) in more detail under different (i) compiler optimizations and (ii) when different security labels are assigned to the \texttt{idx} variable. We summarize our findings as follows.

\paragraph{Case \#1} When \texttt{idx} is labelled as a \textbf{public} variable and the code is compiled with optimization level \optzero{}, \scamv{} detects the leakage and refines the \slh{} hardening. However, all protections are ultimately removed due to the effect of code alignment changes as discussed in~\ref{sec:security:profromance}.
When \texttt{idx} is labelled as \textbf{private}  and the code is compiled with optimization \optzero{}, \scamv{} can also detect the leakage. Subsequently, \scamv{} refines the \slh{} protection to retain only the hardening of the vulnerable load. However, this refinement did not result in a performance improvement.

Similarly, with a \textbf{public} \texttt{idx} and the optimization level set to \opttwo{}, \scamv{} identifies the leakage and refines the \slh{} protection to retain only those that are essential for the vulnerable load. Notably, our refinement did not improve performance in this case either. 
Finally, when \texttt{idx} is labelled as \textbf{private} and the code is compiled with \opttwo{} enabled, \scamv{} detects the leakage. However, the LLVM \slh{} implementation fails to prevent leakage. As a result, we employed alternative hardening methods for the program: \emph{\slh{} on memory addresses} (aSLH\footnote{We borrowed the terminology from ~\cite{zhang2023ultimate}}) and \emph{fence insertion}. Refining aSLH did not improve performance. In the latter case, the speculative barrier formed by \dsb{} and \isb{} increased the execution time.

\paragraph{Case \#10}
When \texttt{idx} is labelled as a \textbf{public} value, and the code is compiled with optimization level \optzero{}, \scamv{}  does not identify any leakage. Thus, there is no need to protect the code with \slh{}.
However, when \texttt{idx} is labelled as \textbf{private} and the code is compiled with optimization \optzero{}, \scamv{} successfully detects the leakage. Subsequently, \scamv{} refines the \slh{} protection to retain only the essential hardening for the vulnerable load. Notably, our \slh{} refinement enhances performance in this scenario.

With a \textbf{public} \texttt{idx} and an optimization level set to \opttwo{}, \scamv{} fails to detect any leakages, making hardening unnecessary. 
Finally, when \texttt{idx} is considered as \textbf{private} and the code is compiled with  \opttwo{} enabled, \scamv{} identifies the leakage. However, \slh{} does not provide effective protection to stop the leakage. Therefore, alternative countermeasures, including refined aSLH and fence insertion, are employed. In this scenario, refining aSLH results in performance improvement.

\subsection{Analyzing crypto libraries}
\label{sec:crypto:analysis}
% \todo[inline]{ what we need here:
%     (1) functions we analyzed;
%     (2) number of well-formed slices;
%     (3) number of detected leakages}
% % For evaluating the \textbf{scalability}, we applied \scamv{} to the codebase of XXX and YYY crypto libraries.
To show \scamv{} scales to real-world case studies, we used Spectre-PHT gadgets in OpenSSL v3.1.0 that had been discussed in \cite{Mosier2022}. Among the three vulnerable gadgets, namely EVP\_PKEY\_asn1\_get0, ts\_check\_status\_info and SSL\_get\_shared\_sigalgs, \scamv{} identifies only SSL\_get\_shared\_sigalgs (due to a leakage at line \#44 in the snippet below) to be vulnerable on actual hardware (see Table~\ref{tab:further-results}). Others did not trigger speculation as the speculation primitive's (i.e., branch condition) operands were already loaded into the cache by the code executed before the \texttt{if} statement.
To protect against SSL\_get\_shared\_sigalgs leakage, we initially applied the default LLVM \slh{} countermeasure. However, employing \slh{} solely on loaded values did not mitigate the leakage. Subsequently, by adopting \slh{} on the memory load addresses, we could stop the leakage.

\begin{figure}[h]
\centering
\includegraphics[]{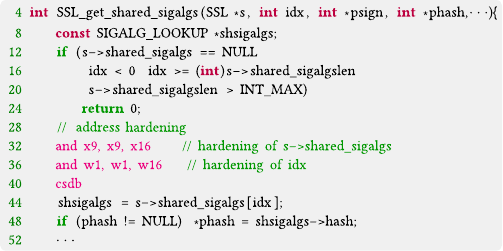}
\Description{Function from OpenSSL v3.1.0 that is vulnerable to Spectre-PHT attack. There is a dangerous memory access in which the miss-speculation could potentially trigger a speculative out-of-bounds load, allowing access to an arbitrary secret.}
\end{figure}

%%% Local Variables:
%%% mode: latex
%%% TeX-master: "../main"
%%% End:

\section{Related work}
\label{sec:related}

Several studies tried to identify and mitigate transient execution attacks. We only discuss a few studies relevant to our results. For a comprehensive list of existing work, we refer the reader to~\cite{survay:transitiveexecattacks,cauligi:sok-spectre-sw-defense}.

\paragraph{\bf Detecting Spectre Attacks}
Widely used techniques for detecting Spectre-style vulnerabilities include \emph{Symbolic execution} and \emph{relation analysis}~\cite{guarnieri:spectector,2021:huntingthehaunter,scamv},  as well as \emph{fuzzing}~\cite{oleksenko:specfuzz,revizor}, both at the machine code ~\cite{2021:oo7,guarnieri:spectector,ctfoundations:pldi:20,2021:huntingthehaunter,oleksenko:specfuzz,formalapproach:csf:19} and LLVM-IR levels~\cite{guanhua:kleespectre,wu:abstractinterpretation,guo:specusym}.
Yet, most existing tools either do not scale well or face qualitative limitations.
For example, SpecFuzz~\cite{oleksenko:specfuzz} simulates the execution of code fragments in misspeculated branches and uses input fuzzing to pinpoint programs' vulnerability to Spectre attacks. However, it does not perform well for nested speculation and inherits limitations of fuzzing, e.g., input coverage. 
Scam-V~\cite{scamv} uses instruction fuzzing and relational testing to synthesize test cases and check the vulnerability of  modern processors to Spectre-PHT. A similar approach was used by Revizor~\cite{revizor,revisor2} to identify Spectre-PHT/STL (Store-to-Load forwarding).
However, in contrast to Revizor, Scam-V utilizes observation refinement and symbolic execution to guide input generation and reduce the search space, thus requiring fewer test cases to uncover the potential leakages~\cite{scamv:buiras2021}. 
The other symbolic execution-based approaches include Spectector~\cite{guarnieri:spectector,DBLP:conf/ccs/FabianGP22} (detects Spectre-PHT/STL/RSB (Return Stack Buffer)), KLEESpec~\cite{guanhua:kleespectre} (detects Spectre-PHT), Pitchfork~\cite{ctfoundations:pldi:20} (detects Spectre-PHT/STL), and BH~\cite{2021:huntingthehaunter} (detects Spectre-PHT/STL); all with scalability limitations. 

\paragraph{\bf Software Mitigations} 
There is also a growing body of work that (formally) analyze programs' vulnerabilities and mitigate leakages using software measures~\cite{barthe:highassurance,ctfoundations:pldi:20,patrignani2021exorcising,guarnieri:spectector,vassena:blade,guanciale:inspectre,guarnieri:contracts,shivakumar:spectredeclassified}. For example, oo7~\cite{2021:oo7} uses taint tracking to find Spectre-PHT attacks and inserts \lfence{} to stop the leakage.
Cauligi et al.~\cite{ctfoundations:pldi:20} proposed Pitchfork based on the concept of speculative constant-time for speculative execution. However, while their theoretical developments suggest inserting fences to mitigate leakages, Pitchfork does not provide this in practice.
InSpectre~\cite{guanciale:inspectre} offers an operational model to aid in reasoning about countermeasures and transient execution attacks.
Patrignani and Guarnieri~\cite{patrignani2021exorcising} analyzed the effects of compiler transformations and countermeasures on speculative execution security. They showed that the existing \slh{} mitigation in LLVM is inadequate for stopping Spectre-PHT leakage in programs and proposed a more powerful version of \slh{} to prevent data leaks.
Shivakumar et al.~\cite{shivakumar:spectredeclassified} demonstrated the ineffectiveness of the LLVM primitives to mitigate Spectre-PHT, proposing a new \slh{} variant to address the limitations of the existing LLVM \slh{} mitigation. 
Blade~\cite{vassena:blade} employs a static type system to detect transient leakage and uses \lfence{s} or \slh{} to mitigate the found leaks in constant-time WebAssembly. None of these mitigations are easily deployable in an existing toolchain, such as LLVM's \lfence{} and \slh{} mitigations.
Mosier et al.~\cite{lcms:mosier:2022} introduced leakage containment models (LCMs), which are \emph{axiomatic} security contracts designed to formally model and automatically detect leakages in programs. 

While some of these works use static analysis techniques to optimize the number of fences, e.g.,~\cite{2021:oo7,lcms:mosier:2022},  we are not aware of any work optimizing fence placement by consulting the hardware. 

\paragraph{\bf Hardware Mitigations}
The research community also proposed several hardware defenses against Spectre attacks. Hardware-level mitigations can be grouped into two main classes.
The first class are techniques that hide the effect of speculative access  instructions~\cite{invisispec,safespec,delayonmiss,cleanupspec} by, e.g., introducing a speculative buffer~\cite{invisispec} or shadow hardware structures to squash microarchitectural state changes if the processor mispredicts~\cite{safespec}. The second class includes techniques that leverage information flow tracking to block leakages by 
preventing data forwarding between speculatively executed access and transmitter instructions~\cite{jiyong:stt,nda,dolma}.

\paragraph{\bf Hardware-Software Co-design} In order to deliver the promised security guarantees without sacrificing performance, hardware-based mitigations require significant modifications in hardware. Instead, there exist also works that propose a software-hardware co-design approach. Examples of such techniques include~\cite{contextsensitivefencing,speccfi,conditionalspec}. For example, Taram et al.~\cite{contextsensitivefencing} proposed the concept of context-sensitive fencing that uses taint tracking to find the optimal location for inserting fences at the decoder level. They also make various speculative barriers available to software. 
\section{Concluding Remarks}
We explored the necessity of hardenings introduced by the LLVM \slh{} pass against Spectre-PHT by taking into account the properties of the underlying microarchitecture.
Our experiments highlighted several interesting results. We showed that the vulnerability of programs to Spectre attacks and the required level of protection to stop potential leaks strictly depend on the properties of the underlying processor and the compiler optimization level. Additionally, we showed that there are unexpected factors (e.g., code alignment) that can impact the vulnerability of programs to side-channel attacks. 

\scamv{'s} current implementation only supports the ARM and RISC-V architectures, but porting it to other architectures like x86 just requires extending the binary-to-BIR translation module. Moreover, we only focused on Spectre-PHT. Covering other variants like Spectre-STL mainly requires developing new observation refinement techniques to synthesize an equivalence relation that can be used to generate suitable test cases and training data w.r.t the variant under test.

%%% Local Variables:
%%% mode: latex
%%% TeX-master: "../main"
%%% End:

\section*{Acknowledgments}
This work was supported in part by a gift from Intel. We thank the anonymous reviewers for their valuable feedback during the review process.

\bibliographystyle{ACM-Reference-Format}
\bibliography{biblio}

\end{document}